\pdfoutput=1
\documentclass[journal]{IEEEtran}
\usepackage{cite}
\usepackage{amsmath,amssymb,amsfonts}
\usepackage{graphicx}
\usepackage{textcomp}
\usepackage{booktabs}
\usepackage{threeparttable}
\usepackage{url}
\usepackage{algorithm}
\usepackage{algpseudocode}
\graphicspath{{figures/}{./}}

\newcommand{\gu}{g_{\mathrm{unif}}}
\newcommand{\B}{\mathcal{B}}
\newcommand{\gbar}{\bar g}

\newtheorem{theorem}{Theorem}
\newtheorem{proposition}{Proposition}
\newtheorem{remark}{Remark}

\makeatletter
\@ifundefined{proof}{\newenvironment{proof}[1][Proof]{\par\noindent\emph{#1.}\ }{\hfill$\square$\par}}{}
\makeatother

\usepackage[hidelinks]{hyperref}
\hypersetup{
  pdftitle={QAOA Inside ADMM: Subproblem Wins Do Not Become Decomposition Wins at Matched Budgets},
  pdfauthor={Mohammad Arif Hossain and Sanim Yousuf Fahim},
  pdfkeywords={QAOA, ADMM, mixed-binary optimization, benchmarking, parameter transfer}
}

\begin{document}

\title{QAOA Inside ADMM: Subproblem Wins Do Not Become Decomposition Wins at Matched Budgets}

\author{Mohammad~Arif~Hossain$^{*}$ and Sanim~Yousuf~Fahim$^{*}$%
\thanks{This work has been submitted to the IEEE for possible publication. Copyright may be transferred without notice, after which this version may no longer be accessible.}%
\thanks{M.~A.~Hossain is with the Department of Engineering Technology \& QRISE Center, Middle Tennessee State University, Murfreesboro, TN 37132 USA (e-mail: \mbox{mohammad.hossain@mtsu.edu}).}%
\thanks{S.~Y.~Fahim is with Daffodil International University, Dhaka, Bangladesh (e-mail: \mbox{fahim23105341092@diu.edu.bd}).}%
\thanks{$^{*}$These authors contributed equally to this work. Corresponding author: Mohammad Arif Hossain.}}

\markboth{Preprint}{Hossain and Fahim: QAOA Inside ADMM}

\maketitle

\begin{abstract}
We benchmark the quantum approximate optimization algorithm (QAOA) as the binary subproblem solver in a two-block alternating direction method of multipliers (ADMM) for mixed-binary problems against classical solvers under matched budgets. With 23 optimizer evaluations per subproblem, scoring determines the comparison: on its final readout, the optimized QAOA loses to random search at $n=12$ and 16 binary variables (inconclusive at 20; exploratory and post hoc tests). When credited with its best sample, it wins. In a noiseless 12-qubit simulation, transferred-angle depth-2 QAOA finds the exact optimum of 97.3\% of planted-family subproblems, against 50.6\%--67.4\% for the four strongest local heuristics (100\% against 90.4\%--94.4\% crediting the four best bitstrings) and a not significantly different 96.5\% for semidefinite rounding with local search (SDP+LS). Under our evaluation-counting convention, its time to solution there is a factor of three to five below the local heuristics' but grows with a larger exponent; it is ten times that of SDP+LS and, in seconds, 54 times that of tabu search with restarts. For the benchmark class, an unconstrained binary quadratic problem in disguise, ADMM's binary step is a proximal step on the reduced problem: at unit penalty ADMM re-solves it, and for larger penalties exact solutions increasingly stall. On two families that do not reduce to one, less accurate solvers found the problem optimum more often at unit and larger penalties. No family gave QAOA a better ADMM outcome than the best classical solver and penalty. The benchmark includes a correctness gate and a log of its hypothesis tests.
\end{abstract}


\section{Introduction}
\label{sec:intro}

\IEEEPARstart{D}{ecomposition} methods are a standard route to large mixed-binary problems, such as unit commitment in power systems \cite{mahroo2022hybrid}, because they split a problem into continuous and combinatorial parts that can be solved separately. The alternating direction method of multipliers (ADMM) \cite{boyd2011distributed} is among the most widely used, also as a heuristic for mixed-integer programs \cite{takapoui2020simple}. Gambella and Simonetto \cite{gambella2020multiblock} showed that its binary update, a quadratic unconstrained binary optimization (QUBO) subproblem, can be delegated to a quantum heuristic once per outer iteration, a role for which the quantum approximate optimization algorithm (QAOA) \cite{farhi2014quantum} is a leading near-term candidate \cite{preskill2018nisq,cerezo2021variational,bharti2022noisy,abbas2024challenges}.

Whether the quantum subroutine is worth its cost is an open question. Benchmarks consistently show that the answer depends on how cost and quality are accounted for \cite{lubinski2024optimization,koch2026quantum,streif2019comparison,pelofske2024short,lykov2023sampling}, and we find one accounting choice to be decisive. A QAOA run draws hundreds of bitstrings while optimizing its angles, yet its output can be the best bitstring of a short final read-out. The Qiskit Optimization tutorial (version 0.7.0) for the ADMM heuristic of \cite{gambella2020multiblock} does this: it builds the QUBO solution from the final sampled state (\texttt{eigenstate}), not from the best bitstring measured during optimization (\texttt{best\_measurement}) \cite{qiskitoptimization,qiskitalgorithms}; with the tutorial's sampler (1,024 shots per circuit call by default), this read-out is a single call's shots. Simulated annealing (SA) and random search, in contrast, are credited with their best evaluation. Under a matched budget, this asymmetry reverses the comparison, as we show.

Without the asymmetry, the question is which QAOA protocol best uses a fixed budget. Per-instance angle optimization spends most of its evaluations searching a noisy parameter landscape, whereas angles trained once on small instances and reused \cite{brandao2018forfixed,akshay2021concentration} make every evaluation a candidate solution; ADMM, whose subproblems change at every outer iteration, is a demanding test of such transferability. Solvers are judged at two levels, kept separate throughout: by the QUBO solutions they return within their budget (\emph{subproblem level}) and by the best feasible solution of the original problem that ADMM finds with them (\emph{ADMM level}). We study the fixed-penalty two-block scheme (Algorithm~\ref{alg:admm}) on problem \eqref{eq:mbo}, an instance of the quadratic-plus-convex class of \cite{gambella2020multiblock} whose continuous variables can be eliminated so that every ADMM run can be checked against the exact optimum.

In every family tested, some classical configuration of solver and penalty found the optimum of the original problem at least as often as the best QAOA configuration. The paper makes four contributions.

\begin{enumerate}
\item \emph{The scoring rule decides the comparison.} With a small training budget of 23 optimizer evaluations per subproblem, optimized QAOA loses to random search under final-read-out scoring and wins under best-sample scoring of the same logged runs, at $n=12$ for every tested ADMM penalty $\rho$ from 0.5 to 4 and at $n=16$ and 20 for $\rho=2$ (inconclusively at $n=20$, uncorrected $P=0.041$; the read-out tests are exploratory and, except at $n=12$ and $\rho=2$, post hoc). By Proposition~\ref{prop:scoring}, read-out scoring requires the final state to amplify every small good set roughly $\B/R$-fold over uniform sampling (budget $\B$, read-out size $R$). This is an accounting caution for this regime, not a QAOA property.

\item \emph{Transferred angles against tuned classical solvers.} Depth-2 transferred-angle QAOA matches seven classical solvers, three of them tuned, on the benign main family. On a Wishart planted family \cite{hamze2020wishart}, it finds the exact optimum more often than tuned tabu search with restarts, tuned parallel tempering (PT), tuned SA, and local search, also when near-degenerate competitors are credited. Its time-to-solution is shorter than theirs by a prefactor in evaluations (charging tabu search and local search $n$ evaluations per neighborhood scan), but it grows with a larger exponent, has no advantage over semidefinite-programming (SDP) rounding with local search (SDP+LS), and is much larger in seconds.

\item \emph{What ADMM's binary step solves.} Theorem~\ref{thm:admm} shows that after the first iteration, the binary subproblem is a proximal step on the reduced problem and characterizes the bitstrings at which exact subproblem solutions stall. A penalty sweep is consistent with every prediction that a finite run can test. In two mixed-binary families outside the theorem, on one of which QAOA was among the most accurate subproblem solvers, less accurate solvers found the problem optimum more often for $\rho\ge1$.

\item \emph{A benchmark protocol with a correctness gate.} The benchmark includes a gate at the matrix, circuit, and sampling levels and a log of its hypothesis tests with cluster-adjusted inference, and we report a circuit-sampler limitation that the gate did not test (Section~\ref{subsec:sampling}).
\end{enumerate}

\begin{figure*}[t!]
\centering
\includegraphics[width=\textwidth]{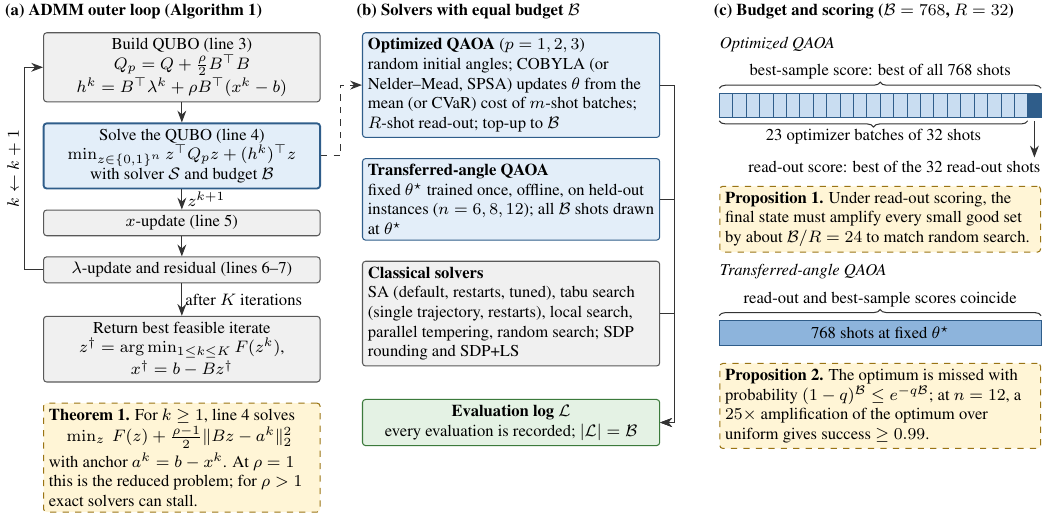}
\caption{Method overview. (a) ADMM for problem \eqref{eq:mbo} delegates its binary update to a QUBO solver once per iteration; by Theorem~\ref{thm:admm}, this QUBO is a proximal step on the reduced problem. (b) Every solver receives the same budget of $\B$ objective evaluations, all logged. (c) How optimized and transferred-angle QAOA spends the budget in the main protocol, and which shots each scoring rule credits.}
\label{fig:method}
\end{figure*}

Fig.~\ref{fig:method} summarizes the method. Section~\ref{sec:related} reviews related work, Sections~\ref{sec:prelim} and~\ref{sec:theory} give the problem and analytical results, Sections~\ref{sec:design} and~\ref{sec:setup} the benchmark, Section~\ref{sec:results} the results, and Sections~\ref{sec:discussion} to~\ref{sec:conclusion} the discussion, limitations, and conclusions. The Appendix gives the proofs.

\section{Related Work}
\label{sec:related}

\subsection{QAOA and ADMM for Mixed-Binary Problems}

Gambella and Simonetto \cite{gambella2020multiblock} introduced multiblock ADMM heuristics that delegate the binary block of quadratic-plus-convex mixed-binary problems to a variational quantum eigensolver \cite{peruzzo2014variational} or to QAOA. Such hybrids have been applied to transaction settlement \cite{braine2021quantum} and unit commitment \cite{mahroo2022hybrid}, but to our knowledge, none compares the quantum subroutine with classical solvers at a matched budget. Over a nonconvex set such as $\{0,1\}^n$, ADMM is a heuristic, used with restarts because it may converge to suboptimal points \cite{takapoui2020simple}. Convergence analyses of nonconvex ADMM require a penalty large relative to the smoothness modulus of the continuous block \cite{wang2019global,li2015global}; Theorem~\ref{thm:admm} uses the same first-order identity to show that, for the binary block, a large penalty enlarges the set of suboptimal stalling points.

\subsection{Benchmarking QAOA Against Classical Solvers}

QAOA has been compared with quantum and simulated annealing \cite{streif2019comparison}, unstructured search \cite{golden2023numerical}, and classical MaxCut heuristics by the samples needed to reach a given quality, a measure that credits QAOA with its best sample \cite{lykov2023sampling}; it has also been calibrated against the classical effort matching its quality \cite{dupont2022calibrating}. Hybrid variational heuristics were far less efficient than classical solvers on small instances \cite{nannicini2019performance}, quantum annealing outperformed short-depth QAOA on 127-qubit higher-order Ising models \cite{pelofske2024short}, and QAOA with amplitude amplification showed a scaling advantage over exact solvers on low-autocorrelation binary sequences \cite{shaydulin2024evidence}. Application-oriented suites weigh solution quality against execution cost \cite{lubinski2024optimization} and provide hard instances \cite{koch2026quantum}. Meaningful scaling exponents require each solver's run length to be optimized at every size \cite{ronnow2014defining,albash2018demonstration}, and stochastic algorithms need explicit budgets and replication \cite{bartzbeielstein2020benchmarking}. To our knowledge, no prior work shows that the choice between read-out and best-sample scoring reverses a matched comparison.

\subsection{Fixed, Concentrated, and Transferred Angles}

At fixed angles, the QAOA objective concentrates over typical instances \cite{brandao2018forfixed}, optimal angles concentrate as size grows \cite{akshay2021concentration}, and fixed angles are near optimal for MaxCut on regular graphs \cite{wurtz2021fixed}. Angles transfer between weighted MaxCut instances \cite{shaydulin2023parameter}, aided by weight rescaling \cite{sureshbabu2024parameter}, which motivates our normalization; structured initializations exploit the same regularities \cite{zhou2020qaoa}. The success-probability decay of fixed-angle QAOA has been compared with classical solvers on random $k$-SAT \cite{boulebnane2024solving} and for linear-ramp schedules \cite{montanezbarrera2025toward}. On MaxCut, local classical algorithms match or exceed the performance of bounded-depth QAOA, which sees only a neighborhood of each variable \cite{farhi2020needs,hastings2019classical,barak2022classical}; we therefore compare the scaling of QAOA with that of local heuristics.

\subsection{Angle Optimization and Classical Baselines}

The classical optimizer strongly affects QAOA \cite{fernandez2022study}: conditional value-at-risk (CVaR) aggregation \cite{barkoutsos2020improving} and warm starts \cite{egger2021warmstarting} are common remedies, whereas shot noise \cite{scriva2024challenges} and barren plateaus \cite{mcclean2018barren} limit any optimizer. Among classical baselines, the Goemans--Williamson (GW) relaxation \cite{goemans1995improved} guarantees an expected ratio of 0.878 for MaxCut with nonnegative weights. Relaxation-based and local-search heuristics, including tabu search, are among the strongest for MaxCut and QUBO \cite{dunning2018what,palubeckis2004multistart}, and parallel tempering \cite{hukushima1996exchange,earl2005parallel} is the standard spin-glass sampler. QAOA has also been run on superconducting processors \cite{harrigan2021quantum}. We compare solvers in noiseless simulation, with a check under depolarizing noise, and use a quantum processor only to measure the device time per shot.

\section{Problem Formulation}
\label{sec:prelim}

\subsection{ADMM for Mixed-Binary Optimization}

We consider an instance of the quadratic-plus-convex mixed-binary class studied in \cite{gambella2020multiblock},
\begin{equation}
\min_{x\in\mathbb{R}^n,\,z\in\{0,1\}^n}\ \tfrac12\|x\|_2^2-c^{\top}x+z^{\top}Qz\quad\text{s.t.}\quad x+Bz=b,
\label{eq:mbo}
\end{equation}
with $Q,B\in\mathbb{R}^{n\times n}$, $Q$ symmetric, and $c,b\in\mathbb{R}^n$. With penalty $\rho>0$ and dual variable $\lambda$, the augmented Lagrangian is $L_\rho(x,z,\lambda)=\tfrac12\|x\|_2^2-c^{\top}x+z^{\top}Qz+\lambda^{\top}(x+Bz-b)+\tfrac{\rho}{2}\|x+Bz-b\|_2^2$. Each iteration updates $z$, then $x$, then $\lambda$ (Algorithm~\ref{alg:admm}). For fixed $(x^k,\lambda^k)$ the $z$-update is the QUBO
\begin{equation}
z^{k+1}\in\arg\min_{z\in\{0,1\}^n} z^{\top}Q_pz+(h^k)^{\top}z,
\label{eq:zupdate}
\end{equation}
\begin{equation}
Q_p=Q+\tfrac{\rho}{2}B^{\top}B,\qquad h^k=B^{\top}\lambda^k+\rho B^{\top}(x^k-b),
\label{eq:zqubo}
\end{equation}
followed by $x^{k+1}=\big(c-\lambda^k-\rho(Bz^{k+1}-b)\big)/(1+\rho)$ and $\lambda^{k+1}=\lambda^k+\rho(x^{k+1}+Bz^{k+1}-b)$, starting from $x^0=\lambda^0=0$. Because $h^k$ changes at every iteration, the solver faces a new QUBO each time. For any $z$, the point $x=b-Bz$ is feasible, so every iterate $z^k$ defines a feasible solution of \eqref{eq:mbo}. Eliminating $x$ gives the reduced objective

\begin{equation}
F(z)=\tfrac12\|b-Bz\|_2^2-c^{\top}(b-Bz)+z^{\top}Qz,
\label{eq:reduced}
\end{equation}
itself a QUBO. Problem \eqref{eq:mbo} is therefore a QUBO in disguise: an exact or heuristic QUBO solver applied to $F$ solves it directly. We study it because it is the class of \cite{gambella2020multiblock} and because enumeration yields its global optimum for the sizes studied, thereby enabling the ADMM-level verification of Section~\ref{subsec:admm}. Theorem~\ref{thm:admm} relates \eqref{eq:zupdate} to \eqref{eq:reduced}. Algorithm~\ref{alg:admm} returns the best feasible iterate, $z^\dagger=\arg\min_{1\le k\le K}F(z^k)$ with $x^\dagger=b-Bz^\dagger$. Section~\ref{subsec:general} also studies two families that do not reduce to a QUBO.

\begin{algorithm}[t]
\caption{ADMM with a QUBO subproblem solver}
\label{alg:admm}
\begin{algorithmic}[1]
\Require $Q,B,c,b$; penalty $\rho$; iterations $K$; subproblem solver $\mathcal{S}$ with budget $\B$
\State $x^0\gets0$, $\lambda^0\gets0$
\For{$k=0,1,\dots,K-1$}
\State $Q_p\gets Q+\tfrac{\rho}{2}B^{\top}B$, $h^k\gets B^{\top}\lambda^k+\rho B^{\top}(x^k-b)$
\State $z^{k+1}\gets\mathcal{S}(Q_p,h^k;\B)$ \Comment{QUBO \eqref{eq:zupdate}}
\State $x^{k+1}\gets\big(c-\lambda^k-\rho(Bz^{k+1}-b)\big)/(1+\rho)$
\State $\lambda^{k+1}\gets\lambda^k+\rho(x^{k+1}+Bz^{k+1}-b)$
\State $\delta^{k+1}\gets\|x^{k+1}+Bz^{k+1}-b\|_2$ \Comment{primal residual}
\EndFor
\State \Return best feasible iterate $(b-Bz^\dagger,z^\dagger)$, $z^\dagger=\arg\min_{1\le k\le K}F(z^k)$
\end{algorithmic}
\end{algorithm}

\subsection{QAOA and the Cost Hamiltonian}
\label{subsec:hamiltonian}
Writing $z_i=(1-s_i)/2$ with $s_i\in\{\pm1\}$ gives the cost $C(z)=z^{\top}Q_pz+h^{\top}z$ as
\begin{align}
C(s)&=\text{const}+\sum_i\Big(-\tfrac{h_i}{2}-\tfrac{Q_{p,ii}}{2}\Big)s_i \nonumber\\
&\quad-\sum_{i<j}\tfrac{Q_{p,ij}+Q_{p,ji}}{4}(s_i+s_j)+\sum_{i<j}\tfrac{Q_{p,ij}+Q_{p,ji}}{4}s_is_j,
\label{eq:ising}
\end{align}
whose operator form, with $s_i\mapsto Z_i$, defines the cost Hamiltonian $H_C$. Every off-diagonal coupling contributes a linear term to both of its spins. An implementation that updates only one accumulator per ordered pair halves these terms; we found this defect in our initial implementation, where the ground state of the defective Hamiltonian differed from the true optimum of the first subproblem on 66 of 80 test instances. QAOA of depth $p$ prepares
\begin{equation}
|\boldsymbol{\gamma},\boldsymbol{\beta}\rangle=U_M(\beta_p)U_C(\gamma_p)\cdots U_M(\beta_1)U_C(\gamma_1)|+\rangle^{\otimes n},
\label{eq:qaoa}
\end{equation}
with $U_C(\gamma)=e^{-i\gamma\tilde H_C}$, $U_M(\beta)=e^{-i\beta H_M}$, and $H_M=\sum_iX_i$, where $\tilde H_C=H_C/\sigma$ and $\sigma$ is the root-mean-square of the non-identity Pauli coefficients of $H_C$ ($\sigma=1$ in the unnormalized variant). The normalization makes angle scales comparable across subproblems with different energy scales \cite{sureshbabu2024parameter}.

\subsection{Normalized Gap}
For a subproblem with optimum $C^*$, mean cost $\bar C$ over all $2^n$ bitstrings, and maximum $C_{\max}$, the normalized gap of a bitstring $z$ is
\begin{equation}
\gu(z)=\frac{C(z)-C^*}{\bar C-C^*},
\label{eq:gu}
\end{equation}
which is 0 at the optimum, 1 for the expected result of a single uniform guess, and at most $G=(C_{\max}-C^*)/(\bar C-C^*)$. For a distribution over bitstrings, $\gbar$ denotes its mean normalized gap, so that $\gbar=1$ for the uniform distribution.

\section{Analytical Results}
\label{sec:theory}

This section states three results in terms of Algorithms~\ref{alg:admm} and~\ref{alg:qaoa}; Section~\ref{sec:results} tests each, and the Appendix gives the proofs. Propositions~\ref{prop:scoring} and~\ref{prop:fixed} are elementary consequences of independent sampling; we state them because they set the quantitative bars that the experiments test.

\subsection{Scoring Rules and Sampling Budgets}
Call a set $\Omega$ of bitstrings \emph{good} (for example, all bitstrings whose normalized gap is at most some threshold), and let $u$ be the fraction of all $2^n$ bitstrings that lie in $\Omega$. Random search with budget $\B$ finds $\Omega$ with probability $1-(1-u)^{\B}$. We ask what the final QAOA state must achieve to do as well when only its read-out is scored.

\begin{proposition}[Read-out scoring raises the bar]
\label{prop:scoring}
Let the QAOA solver of Algorithm~\ref{alg:qaoa} use budget $\B$ with a read-out of $R$ shots drawn independently from its final state, and let that state place probability $\pi$ on a good set $\Omega$. Then:
\begin{enumerate}
\item[(a)] the best-sample output is never worse than the read-out output;
\item[(b)] the read-out finds $\Omega$ at least as often as random search if and only if
\begin{equation}
\pi\ \ge\ 1-(1-u)^{\B/R}\ \approx\ \frac{\B}{R}\,u,
\label{eq:readout}
\end{equation}
where the approximation holds when $(\B/R)\,u$ is small, and the exact threshold always lies between $0.63\,(\B/R)\,u$ and $(\B/R)\,u$ once $(\B/R)\,u\le1$;
\item[(c)] if all $\B$ shots are drawn from one state and the best sample is scored, as with transferred angles, the condition is only $\pi\ge u$.
\end{enumerate}
\end{proposition}

In words, read-out scoring requires the final state to amplify every small good set by roughly $\B/R$ over uniform sampling (larger sets by the smaller factor of the exact threshold), whereas best-sample scoring with fixed angles requires no amplification to match random search. In the main protocol, $\B/R=768/32=24$. Because the expected gap equals $\int_0^G\Pr[\text{miss }\Omega_t]\,dt$ with $\Omega_t=\{z:\gu(z)\le t\}$, the mean-gap comparison integrates \eqref{eq:readout} over all thresholds. The per-set condition is therefore sufficient but not necessary for matching random search in the mean gap, and a state can pass it for the optimum yet lose on the mean gap.

\begin{proposition}[Success of fixed-angle sampling]
\label{prop:fixed}
If a fixed state places probability $q$ on the optimal bitstrings and is sampled $\B$ times independently, the optimum is missed with probability
\begin{equation}
(1-q)^{\B}\ \le\ e^{-q\B},
\label{eq:miss}
\end{equation}
and the expected normalized gap of the best sample is at most $G\,(1-q)^{\B}$. Hence $q\ge\ln(1/\epsilon)/\B$ guarantees that the optimum is found with probability at least $1-\epsilon$.
\end{proposition}

For uniform sampling at $n=12$ with a unique optimum, $q=2^{-12}$ and $\B=768$ give a success probability of $1-(1-2^{-12})^{768}=0.171$. Success with probability 0.99 requires $q\ge6.0\times10^{-3}$, an amplification of the optimum by a factor of about 25 over uniform sampling. At a fixed budget the required amplification doubles with every added variable, so Proposition~\ref{prop:fixed} turns the scaling question into how fast $q$ decays with $n$ (Section~\ref{subsec:fixedbudget}).

\subsection{What the Binary Step of ADMM Solves}
Theorem~\ref{thm:admm} concerns problem \eqref{eq:mbo} as stated: a continuous part $\tfrac12\|x\|^2-c^{\top}x$ and the coupling $x+Bz=b$, with no further constraints on $x$.

\begin{theorem}[The binary step is a proximal step]
\label{thm:admm}
Run Algorithm~\ref{alg:admm} on problem \eqref{eq:mbo} with any subproblem solver and penalty $\rho$. For every iteration $k\ge1$, the QUBO passed to the solver in line~4 is, up to an additive constant,
\begin{equation}
\min_{z\in\{0,1\}^n}\ F(z)+\frac{\rho-1}{2}\,\big\|Bz-a^k\big\|_2^2,\qquad a^k=b-x^k,
\label{eq:prox}
\end{equation}
that is, the reduced problem \eqref{eq:reduced} plus a proximal term that pulls $Bz$ toward the anchor $a^k$ set by the current continuous iterate. Consequently:
\begin{enumerate}
\item[(a)] \emph{Penalty one.} If $\rho=1$, line~4 solves the reduced problem itself at every iteration after the first, so an exact solver finds a global optimum of \eqref{eq:mbo} at the second iteration.
\item[(b)] \emph{Residual.} If line~4 returns the same bitstring $\bar z$ from some iteration on, the residual of line~7 is divided by exactly $1+\rho$ at every subsequent iteration, whether or not $\bar z$ is optimal.
\item[(c)] \emph{Stalling points.} Suppose line~4 is solved exactly, with ties broken in favor of the previous bitstring, and returns $z^{k+1}=\bar z$, computed with the anchor $a^k$, at some iteration $k\ge1$. Then line~4 returns $\bar z$ at every later iteration if and only if
\begin{equation}
F(\bar z)\ \le\ F(z)+\frac{\rho-1}{2}\,\|B(z-\bar z)\|_2^2\quad\forall z\in\{0,1\}^n.
\label{eq:fp}
\end{equation}
We call such a $\bar z$ a \emph{stalling point}; the characterization holds for every $\rho>0$. At $\rho=1$, the stalling points are exactly the global minimizers of $F$. For $\rho>1$ every global minimizer is a stalling point, but suboptimal bitstrings can be as well; their excess cost is at most $\tfrac{\rho-1}{2}\|B(z^\star-\bar z)\|_2^2$ for a global minimizer $z^\star$, and the set of stalling points can only grow as $\rho$ increases. For $\rho<1$ a stalling point must be a global minimizer that is \emph{sharp}, $F(z)-F(\bar z)\ge\tfrac{1-\rho}{2}\|B(z-\bar z)\|_2^2$ for all $z$, and there may be none.
\end{enumerate}
\end{theorem}

For $\rho<1$, the weight in \eqref{eq:prox} is negative; we keep the name \emph{proximal} for brevity. An experiment with $K$ iterations can only observe a bitstring held for its last few iterations, so Section~\ref{subsec:penalty} checks \eqref{eq:fp} directly. Our exact solver breaks exact ties by enumeration order, an event of probability zero for the instance distributions used here.

Theorem~\ref{thm:admm} has three consequences for QAOA-in-ADMM benchmarks. The binary subproblem is a proximally regularized version of the full reduced QUBO, not a smaller problem. A converged residual certifies only that the binary iterate has stopped changing, not that it is good. For $\rho>1$, an exact subproblem solver never leaves a stalling point once it returns one, whereas a stochastic solver keeps changing $z$ and may escape. The first step of the proof is the standard first-order identity of nonconvex ADMM analyses \cite{li2015global}; the contribution lies in the characterization of stalling points in part (c) and its dependence on $\rho$.

\begin{remark}[A general quadratic]\label{rem:general}
The proximal form extends to a strongly convex quadratic $\tfrac12x^{\top}Hx-c^{\top}x$: for $k\ge1$ the QUBO of line~4 equals $F(z)+\tfrac12\|Bz-a^k\|^2_{\rho I-H}$ up to a constant, where $F$ is the corresponding reduced objective and $\|v\|^2_W=v^{\top}Wv$ is a quadratic form, not a norm when $W$ is indefinite. The anchor now contracts toward $B\bar z$ by the matrix $(H+\rho I)^{-1}H$ rather than by the scalar $1/(1+\rho)$. The residual ratio of part (b) therefore needs not be $1+\rho$, and of part (c) only the ``only if'' direction is proved to carry over, with the weight $\rho I-H$ in place of $(\rho-1)I$. Part (a) carries over when $H=\eta I$, at $\rho=\eta$ (Appendix).
\end{remark}

Section~\ref{subsec:general} tests two families outside both results.

\section{Benchmark Design}
\label{sec:design}

\subsection{Matched Objective-Evaluation Budget and Scoring}
One evaluation is one candidate bitstring whose cost the solver obtains: a measurement shot for QAOA, a proposal for SA, a draw for random search, and a hyperplane rounding for the SDP. Tabu search and local search are charged $n$ evaluations per scan of the $n$ one-flip neighbors, although an incremental implementation computes the $n$ flip differences in $O(n)$ time. Every solver receives the same budget $\B$ per subproblem. Because ratios of evaluation counts depend on this convention, we also report wall-clock time (Section~\ref{subsec:tts}); hardware execution time, compilation, and classical optimizer overhead are not equalized.

We log every shot that QAOA draws, in both angle optimization and readout. \emph{Best-sample} scoring credits QAOA with the best bitstring sampled anywhere within its budget, exactly as SA and random search are credited with the best of their evaluations; \emph{read-out} scoring credits QAOA only with the best bitstring of its final read-out, the rule of the Qiskit Optimization path (Section~\ref{sec:intro}). Best-sample scoring is the primary choice because it defines every solver's output consistently. Inside ADMM, the best-sample output updates the iterates, and the read-out is scored on the same trajectory, so the two rules are compared at the subproblem level; the ADMM-level effect of read-out scoring was not measured. If the angle optimizer terminates early, the remainder is spent sampling at the final angles (line~11 of Algorithm~\ref{alg:qaoa}), so every solver uses exactly $\B$ evaluations.

\subsection{Solvers}
\label{subsec:solvers}
\begin{algorithm}[t]
\caption{QAOA subproblem solver with shot logging}
\label{alg:qaoa}
\begin{algorithmic}[1]
\Require QUBO $(Q_p,h)$; budget $\B$; depth $p$; mode; read-out size $R$; shots per evaluation $m$; fixed angles $\theta^\star$
\State build $\tilde H_C=H_C/\sigma$ from \eqref{eq:ising}; $\mathcal{L}\gets\emptyset$ \Comment{log of all shots}
\If{mode is \emph{transferred}}
\State draw $\B$ shots at $\theta^\star$ into $\mathcal{L}$; $\mathcal{R}\gets\mathcal{L}$
\Else \Comment{mode is \emph{optimized}}
\State $\theta\gets$ uniform sample from $[-\pi/4,\pi/4]^{2p}$
\While{optimizer runs and $|\mathcal{L}|+m\le\B-R$}
\State draw $m$ shots at $\theta$ into $\mathcal{L}$
\State return their mean cost to the optimizer; update $\theta$
\EndWhile
\State draw $R$ read-out shots $\mathcal{R}$ at $\theta$ into $\mathcal{L}$
\State draw $\B-|\mathcal{L}|$ shots at $\theta$ into $\mathcal{L}$ \Comment{top-up}
\EndIf
\State \Return best-sample output $\arg\min_{z\in\mathcal{L}}C(z)$ and read-out output $\arg\min_{z\in\mathcal{R}}C(z)$
\end{algorithmic}
\end{algorithm}

\emph{Optimized QAOA} runs constrained optimization by linear approximation (COBYLA) \cite{powell1994direct} (SciPy; initial step 1.0, tolerance $10^{-4}$) from initial angles uniform on $[-\pi/4,\pi/4]$: 23 optimizer evaluations of 32 shots, each returning the mean sampled cost, then a 32-shot read-out at the returned angles ($\B=768$, $R=32$). The main comparison (Fig.~\ref{fig:main}) adds depths $p=1$ and 3, the unnormalized Hamiltonian, warm starts from the angles optimized at the previous ADMM iteration, a multi-start variant that splits the budget over four restarts (each 11 evaluations of 16 shots and a 16-shot read-out) and pools their read-outs, and a budget control that uses a quarter of the budget (11 evaluations of 16 shots and a 16-shot read-out).

\emph{Transferred-angle QAOA} fixes $(\boldsymbol\gamma,\boldsymbol\beta)$ in advance and spends the entire budget sampling one circuit, so its read-out and best-sample scores coincide. Angles are trained on held-out instances of size $n\in\{6,8,12\}$ at $\rho=2$ by minimizing the mean exact expected normalized gap $\gbar$ over 36 subproblems (the first, third, and fifth of the exact ADMM trajectories of 12 instances), with Nelder--Mead \cite{nelder1965simplex} from one linear-ramp and five random starting points. Planted-family angles are trained likewise on held-out planted instances at $n=12$, and the time-to-solution study also uses angles retrained at $n=16$ on 18 subproblems of six held-out instances, starting from the $n=12$ angles and two random points. Training, a one-time classical cost, is excluded from the matched budget, as is the tuning of the classical solvers.

\emph{Classical baselines.} SA \cite{kirkpatrick1983optimization} performs $\B$ single-bit-flip Metropolis proposals with geometric cooling from $T_0=2$ to $0.01$; \emph{SA with restarts} splits the budget over four runs, and \emph{tuned SA} and \emph{SA, tuned (ext.)} use the configurations with the lowest held-out mean gap on grids of 36 and 125 configurations; on the planted family, tuned SA denotes the extended-grid configuration, which is also selected on a wider grid of 100 configurations also selected. \emph{Tabu search} flips the best non-tabu bit along a single trajectory, \emph{local search} runs steepest descent with random restarts, and random search draws $\B$ uniform bitstrings. The three stronger heuristics, charged by the same convention, are \emph{tabu search with restarts} \cite{palubeckis2004multistart,glover1989tabu}, which restarts from a random bitstring when its best value stops improving; \emph{parallel tempering} (PT) \cite{hukushima1996exchange,earl2005parallel} with $M$ replicas, neighbor exchanges after every sweep, and one evaluation per proposal; and \emph{SDP+LS}, which follows each hyperplane rounding of the relaxation below by steepest descent. Tuned solvers were tuned per family on that family's held-out subproblems; on the main family, many configurations tie at a held-out mean gap of zero. Ties went to the first configuration in grid order. For the main and planted families, the selected SA has initial temperature, final temperature, and restarts $(0.25,0.005,16)$ and $(0.5,0.005,16)$ (small grid: $(1,0.1,4)$); tabu search has tenure 3 and 9 and restarts after 6 and 12 non-improving steps; and PT has 8 and 16 replicas on temperature ranges $[0.005,3]$ and $[0.02,0.3]$.

For the ADMM subproblems, a general-QUBO SDP relaxation (linear terms absorbed into couplings to an auxiliary spin) is solved anew for every subproblem with CVXPY \cite{diamond2016cvxpy} and the splitting conic solver (SCS) and followed by $\B$ randomized hyperplane roundings; for couplings of mixed sign, it carries no approximation guarantee. Its solve time is reported, but, like other classical overhead, it is not counted against the evaluation budget.

\subsection{Sampling of QAOA Shots}
\label{subsec:sampling}
QAOA shots are drawn by one of two samplers. The \emph{exact-distribution sampler} computes the output distribution by statevector simulation and draws from it with a generator that persists across calls, so every call returns fresh, independent shots. The \emph{circuit sampler} runs the circuit on the Qiskit Aer sampler, which our code created once per subproblem with a fixed seed and which restarted its seeded stream at every call, so every optimizer evaluation and the read-out of a run reused the same uniform random numbers. Shots at nearby angles are therefore correlated, and the 32-shot read-out at the returned angles, which COBYLA selects for their low sampled mean, repeats exactly the shots of the optimizer evaluation there, as a rerun of the first subproblem of five test instances confirmed. Both effects favor read-out over best-sample scoring and thus work against the reversal of Section~\ref{subsec:scoring}. The circuit sampler drew the optimized-QAOA arms of the main comparison at $\rho=2$ (Fig.~\ref{fig:main}), the 40-instance shared-trajectory comparison, and the budget sweep of Section~\ref{subsec:scoring}. It also drew studies of optimized QAOA with other optimizers (Nelder--Mead and simultaneous perturbation stochastic approximation \cite{spall1998implementation}), CVaR aggregation \cite{barkoutsos2020improving}, and linear-ramp initialization \cite{zhou2020qaoa}, at other sizes, coupling densities, and noise levels, and on MaxCut instances; because common random numbers across calls change the noise that each optimizer sees, we draw no conclusions from these and do not report them. Transferred-angle QAOA, which draws its budget in one call, and every other experiment is unaffected. The penalty sweep repeats the main optimized-QAOA arm on the same instances and seeds with the exact-distribution sampler, with identical classical arms, so at $\rho=2$ it is an independent-draw replicate. Every solver applied to a subproblem receives the same integer seed, which ensures that the arms' pseudorandom draws are correlated without evident bias.

\subsection{Performance Measures}
The primary subproblem-level measure is the normalized gap \eqref{eq:gu} of the solver's output; we also report the fraction of subproblems solved exactly, $\Pr[\text{opt}]$. Exhaustive enumeration provides $C^*$ and $\bar C$ for every subproblem up to $n=24$. At the ADMM level, we report the final primal residual, whether the best feasible iterate is a global optimum of \eqref{eq:mbo} (\emph{problem optimum found}), and the gap between the best feasible objective and that optimum, found by enumerating $z$ in \eqref{eq:reduced} and normalized as in \eqref{eq:gu} with $F$ in place of $C$.

Scaling is measured by the time to solution $\mathrm{TTS}_{99}$, the number of evaluations needed to find the optimum with probability 0.99, with every solver's run length optimized at every size \cite{ronnow2014defining,albash2018demonstration}. For a sampler placing probability $q$ on the optimum (transferred QAOA with exact $q$, uniform sampling, and SDP rounding with $q$ estimated from 4,096 roundings), $\mathrm{TTS}_{99}=\ln0.01/\ln(1-q)$. For solvers with a run length, the run budget $b^\star$ is chosen per solver, size, and family from 25 values between 8 and 32,768 as the one minimizing the median $\mathrm{TTS}_{99}$ over 20 held-out instances (256 runs per budget). Each test instance is then scored from 1,024 runs at $b^\star$ as $b^\star\ln0.01/\ln(1-\hat p)$, with $\hat p$ the fraction of runs that found the optimum, or as $b^\star$, an upper bound, when every run succeeds; an instance with no successful run is censored. The exponent $\kappa$ is the least-squares slope of $\log_2$ of the median against $n$, with a bootstrap interval over instances; exponents are compared by a paired bootstrap of the slope difference (4,000 resamples of the test instances). Conversion to seconds uses the measured serial time per evaluation of our Python implementations, with the SDP relaxation charged once per subproblem, and the device time per shot reported by the runtime service at $n=12$, which is a lower bound for larger $n$.

\subsection{Experimental Designs and Statistics}
\label{subsec:stats}
\emph{ADMM-embedded} runs execute $K=12$ outer iterations unless stated otherwise, with each solver driving its own trajectory; a seed's score is the mean over iterations, and comparisons are paired by instance. \emph{Shared-trajectory} runs score every solver on the identical subproblems of the exact-solver trajectory. \emph{Standalone} runs score the first subproblem of each instance. \emph{Fixed-budget} runs score the second subproblem, the first to take the proximal form of Theorem~\ref{thm:admm}, at $\B=768$ for $n$ from 12 to 24. \emph{Direct} runs apply a solver once to $F$ with the total budget $K\B$ of an ADMM run.

The unit of analysis is the instance. We write $P$ for $P$-values, $r$ for the matched-pairs rank-biserial effect size of the exact two-sided Wilcoxon signed-rank test \cite{wilcoxon1945individual}, with $r<0$ meaning the first-named method is better; because accurate solvers produce many ties, we report the number of non-tied pairs, which bounds the smallest attainable $P$. Intervals are percentile bootstrap 95\% intervals \cite{efron1979bootstrap}. Because consecutive subproblems of an instance are dependent, the fraction solved exactly is compared at the instance level by the cluster-adjusted McNemar statistic of Obuchowski \cite{obuchowski1998comparison} ($P\ge5\times10^{-6}$ from $2\times10^5$ sign flips beyond 20 discordant instances) and, for equivalence, by two one-sided tests with a margin of 2 percentage points. A lenient secondary criterion, a margin of 0.01 in mean $\gu$, places a method among the strong solvers without ranking it. Holm correction \cite{holm1979simple} is applied within families of comparisons.

The planned comparisons are recorded in a log of 392 tests, with a global Bonferroni threshold computed across all tests; equivalence tests, descriptive McNemar counts, and post hoc comparisons are outside the log. A test is labeled confirmatory when its hypothesis was specified before its experiment was run; only the main-family hypotheses preceded all results, and every time-to-solution test is exploratory. The main comparison was extended from 40 to 80 seeds before its results were analyzed, and no experiment was extended in response to the results. Comparisons outside the log, including the post hoc analyses, are identified as such.

\subsection{Correctness Verification}
\label{subsec:gate}
Before any experiment, a gate checks the cost Hamiltonian at three levels and halts on failure. At the matrix level, the diagonal of $H_C$ plus its offset must equal the exact QUBO cost of every bitstring ($n\in\{4,6,8,12,16\}$). At the circuit level ($n\in\{4,8,12\}$), the cost layer with zero mixer angle must prepare $e^{-i\gamma\tilde H_C}|+\rangle^{\otimes n}$, and the full $p=2$ ansatz must reproduce the output probabilities of the independent NumPy simulator used for angle training. At the sampling level ($n=8$, 40,000 shots), the mean sampled cost must agree with the exact expectation within five standard errors, for both samplers. This check examines a single call and does not test independence across calls, which is why the shared random stream of Section~\ref{subsec:sampling} went undetected.

\section{Experimental Setup}
\label{sec:setup}

ADMM instances draw all entries of $Q$ uniformly from $[-1,1]$ (symmetrized), $B$ from $[-0.7,0.7]$, and $c,b$ from $[-1,1]$, and use $\rho=2$ unless stated otherwise. The planted family takes $Q$ from the Wishart planted ensemble \cite{hamze2020wishart}: Ising couplings $J=-WW^{\top}/n$ with zero diagonal, where the $\alpha n$ Gaussian columns of $W$ ($\alpha=0.5$) are orthogonal to a planted configuration hidden by a random gauge, mapped to a QUBO scaled to unit maximum magnitude and weakly coupled to the continuous block ($B$ from $[-0.05,0.05]$, $c,b$ from $[-0.1,0.1]$) so that the subproblems inherit the ensemble's landscape. Because the Wishart Ising model has no field, only these weak linear terms break its symmetry under complementing every bit, and the complement of each optimum is close to it in cost. The main setting is $n=12$ with $\B=768$, 18.75\% of the search space, which keeps exhaustive verification feasible. Test instances use seeds 0 to 79 (planted: 30000 to 30079); angle training and solver tuning use held-out seeds 5000 to 5011, and the selection of run budgets uses 6000 to 6019 (planted: offset by 30000 in both cases). Simulations use Qiskit 2.5.2 \cite{javadi2024quantum}, Qiskit Aer 0.17.2, NumPy \cite{harris2020array}, and SciPy \cite{virtanen2020scipy}. To convert evaluations to seconds, we executed $p=2$ circuits of this benchmark at $n=12$ (4,096 shots each) on the IBM Quantum processor \texttt{ibm\_marrakesh}, for which Qiskit Runtime reported 54~s of quantum processing time for the 163,840 shots of a first submission and 73~s for the 204,800 shots of the five jobs of a second submission that reported usage (its sixth job reported none), both including runs with and without dynamical decoupling and twirling, that is, $3.30\times10^{-4}$ and $3.56\times10^{-4}$~s per shot. We use these executions only for this timing and do not analyze their outputs here (the transpiled circuits contain 526 two-qubit gates, compared with 264 in the logical circuit).

All random number generators are seeded. Our code caches every per-instance result, and each execution after the one that computed it reloads it unchanged, using the same pinned Qiskit versions; all reported results come from executions on Kaggle. Classical timings were obtained in Kaggle's CPU environment using serial Python implementations; the processor model was not recorded. Post hoc analyses, identified as such, were computed from the cached results without new experiments, except for a rerun of five subproblems that checked the circuit sampler (Section~\ref{subsec:sampling}).

\section{Results}
\label{sec:results}

Table~\ref{tab:glance} summarizes the findings.

\begin{table*}[t]
\centering
\begin{threeparttable}
\caption{Findings at a glance.}
\label{tab:glance}
\footnotesize
\setlength{\tabcolsep}{4pt}
\begin{tabular}{@{}p{0.25\textwidth}p{0.72\textwidth}@{}}
\toprule
Question & Finding \\
\midrule
Does the scoring rule matter? & Yes: with 23 optimizer evaluations, optimized QAOA loses to random search under read-out scoring and wins under best-sample scoring of the same runs at $n=12$ and 16 (inconclusive at $n=20$; exploratory or post hoc) (Section~\ref{subsec:scoring}). \\
Do transferred angles solve the subproblems of the main family? & Yes: 99.8\% to 100\% solved exactly with angles trained at $n=6$ to 12, equivalent within 2 percentage points to seven classical solvers; at a fixed budget, $q$ decays as $2^{-0.33n}$ up to $n=24$ at $p=2$ (Sections~\ref{subsec:transfer} to~\ref{subsec:fixedbudget}). \\
Is there an advantage on a rugged planted family? & Over local heuristics only (noiseless, $p=2$): 97.3\% against at most 67.4\% solved exactly; fewer evaluations to solution at $n=12$ under our convention, but a larger exponent; none over SDP+LS or in seconds (Sections~\ref{subsec:hard} and~\ref{subsec:tts}). \\
Does subproblem accuracy improve the ADMM outcome? & No: at $\rho=1$ ADMM re-solves $F$; for $\rho>1$, exact solutions stall more often; in two non-QUBO families, less accurate solvers found the problem optimum more often for $\rho\ge1$ (Sections~\ref{subsec:admm} and~\ref{subsec:general}). \\
\bottomrule
\end{tabular}
\end{threeparttable}
\end{table*}

\subsection{The Scoring Rule Decides the Comparison}
\label{subsec:scoring}

\begin{figure*}[t!]
\centering
\includegraphics[width=\textwidth]{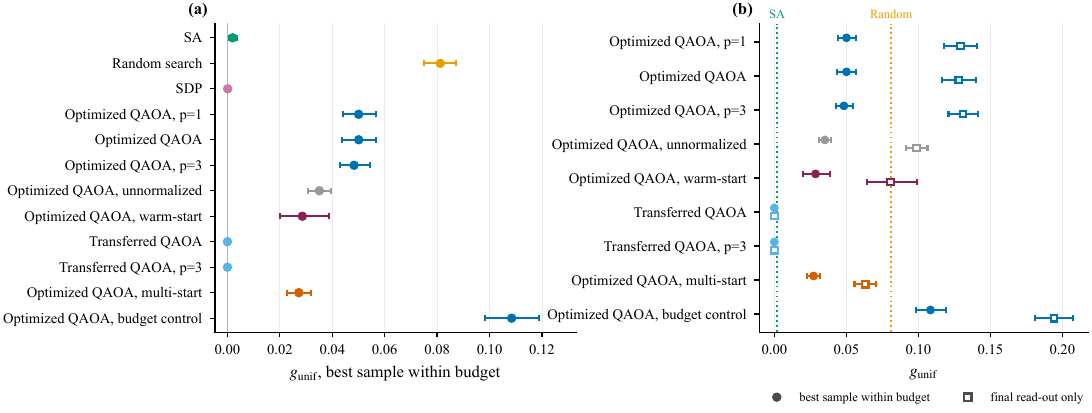}
\caption{ADMM-embedded comparison ($n=12$, 80 instances, $\rho=2$, circuit sampler). (a) Normalized gap $\gu$ under best-sample scoring, mean, and 95\% bootstrap interval. (b) Each QAOA protocol under best-sample (filled) and final-read-out scoring (open); dotted lines: SA and random search. ``Optimized QAOA'' without a depth label is $p=2$. Under read-out scoring, most optimized protocols fall to or behind random search; Table~\ref{tab:crn} gives independent-draw values for optimized QAOA with $p=2$.}
\label{fig:main}
\end{figure*}

Fig.~\ref{fig:main} reports the ADMM-embedded comparison on 80 instances at $\rho=2$, with QAOA drawn by the circuit sampler. Scored on its final read-out, optimized QAOA ($p=2$) is significantly worse than random search ($r=+0.859$ [0.72, 0.96], $P<10^{-10}$); scored on the best bitstring sampled within the same budget, on the same runs, it is significantly better ($r=-0.833$ [$-0.96$, $-0.68$], $P<10^{-10}$, all 80 pairs non-tied) and solves 37.5 percentage points more subproblems exactly (95\% CI 32.7 to 42.1). The independent-draw replicate (Section~\ref{subsec:sampling}) shows this more clearly: the read-out loses ($r=+0.985$, $P<10^{-10}$) and the best sample wins ($r=-0.972$, $P<10^{-10}$; 46.4 percentage points more subproblems solved exactly, 95\% CI 42.0 to 50.6), and the same holds at every penalty from 0.5 to 4 ($P<10^{-10}$ for each; Table~\ref{tab:crn}; post hoc and outside the test log). The unnormalized protocol reverses in the same way but more weakly ($r=+0.42$, $P=9.5\times10^{-4}$), whereas multi-start, whose four pooled 16-shot read-outs halve $\B/R$ to 12, beats random search even under read-out scoring ($r=-0.50$, $P=6.4\times10^{-5}$), as Proposition~\ref{prop:scoring} allows, although the circuit sampler's bias toward the read-out (Section~\ref{subsec:sampling}) may inflate this result. Depth ($p=1,2,3$) had no detectable effect (Friedman test \cite{friedman1937use}, $P=0.99$), a null result that the circuit sampler makes inconclusive. Under best-sample scoring every QAOA variant except the budget control, which uses a quarter of the budget, beats random search, despite the circuit sampler's bias against that rule. Only 32 of the 768 evaluations per subproblem enter the read-out, so read-out scoring discards 96\% of the samples that the budget pays for.

Proposition~\ref{prop:scoring} explains how read-out scoring can lose on the gap while the final state is far from uniform. With independent draws, as the proposition assumes, the read-out of optimized QAOA contained the optimum more often (21.4\% of subproblems) than random search finds it (0.171 in theory, 17.7\% observed), so for the optimal set it met, on average over subproblems, the bar that condition \eqref{eq:readout} sets for each final state. Yet its read-out gap was larger than random search's (Table~\ref{tab:crn}), so for good sets defined by larger gap thresholds, the final states fell short of that bar.

\begin{table*}[t]
\centering
\begin{threeparttable}
\caption{Optimized QAOA ($p=2$) against random search under the two scoring rules, ADMM-embedded. $\Pr[\text{opt}]$: fraction of subproblems solved exactly; $\gu$: mean normalized gap; each for QAOA under best-sample (BS) and read-out (RO) scoring and for random search; $r$: rank-biserial effect of QAOA versus random search ($r>0$: QAOA worse), with exact signed-rank $P$ (all pairs non-tied). Rows 2 to 7: independent-draw replicate on the instances of row 1; last two rows: 15 instances, eight iterations, fixed budget of 768 evaluations.}
\label{tab:crn}
\footnotesize
\setlength{\tabcolsep}{3pt}
\begin{tabular}{@{}llccccccrr@{}}
\toprule
 & & \multicolumn{3}{c}{$\Pr[\text{opt}]$} & \multicolumn{3}{c}{$\gu$} & \multicolumn{2}{c}{$r$ ($P$)} \\
\cmidrule(lr){3-5}\cmidrule(lr){6-8}\cmidrule(l){9-10}
Setting & Sampler & BS & RO & Rand. & BS & RO & Rand. & RO & BS \\
\midrule
$n=12$, $\rho=2$ & circuit & 0.552 & 0.294 & 0.177 & 0.050 & 0.128 & 0.081 & $+0.859$ ($<10^{-10}$) & $-0.833$ ($<10^{-10}$) \\
$n=12$, $\rho=0.5$ & independent & 0.571 & 0.167 & 0.168 & 0.033 & 0.163 & 0.078 & $+0.993$ ($<10^{-10}$) & $-0.994$ ($<10^{-10}$) \\
$n=12$, $\rho=1$ & independent & 0.601 & 0.196 & 0.166 & 0.033 & 0.158 & 0.079 & $+0.965$ ($<10^{-10}$) & $-0.982$ ($<10^{-10}$) \\
$n=12$, $\rho=1.5$ & independent & 0.626 & 0.226 & 0.173 & 0.035 & 0.170 & 0.081 & $+0.982$ ($<10^{-10}$) & $-0.976$ ($<10^{-10}$) \\
$n=12$, $\rho=2$ & independent & 0.641 & 0.214 & 0.177 & 0.037 & 0.170 & 0.081 & $+0.985$ ($<10^{-10}$) & $-0.972$ ($<10^{-10}$) \\
$n=12$, $\rho=3$ & independent & 0.630 & 0.273 & 0.150 & 0.037 & 0.157 & 0.084 & $+0.952$ ($<10^{-10}$) & $-0.983$ ($<10^{-10}$) \\
$n=12$, $\rho=4$ & independent & 0.649 & 0.247 & 0.148 & 0.035 & 0.157 & 0.088 & $+0.966$ ($<10^{-10}$) & $-0.999$ ($<10^{-10}$) \\
$n=16$, $\rho=2$ & independent & 0.333 & 0.075 & 0.017 & 0.078 & 0.229 & 0.150 & $+0.917$ ($6.1\times10^{-4}$) & $-1.000$ ($6.1\times10^{-5}$) \\
$n=20$, $\rho=2$ & independent & 0.142 & 0.008 & 0.008 & 0.114 & 0.258 & 0.207 & $+0.600$ (0.041) & $-1.000$ ($6.1\times10^{-5}$) \\
\bottomrule
\end{tabular}
\end{threeparttable}
\end{table*}

Differing ADMM trajectories do not produce the reversal: on identical subproblems from the exact-solver trajectory (40 instances), optimized QAOA beats random search under best-sample scoring ($r=-0.898$, $P=2.1\times10^{-8}$) and loses under read-out scoring (mean $\gu$ 0.127 against 0.093, $r=+0.72$, $P=2.6\times10^{-5}$, post hoc). Embedded in ADMM at $n=16$ and 20 with a fixed budget of 768 evaluations (15 instances, eight ADMM iterations, independent draws, 1.2\% and 0.07\% of the search space), the read-out had a larger gap than random search at $n=16$ ($r=+0.917$, $P=6.1\times10^{-4}$) and, inconclusively, at $n=20$ ($r=+0.600$, $P=0.041$; both solved 0.8\% of subproblems exactly), and the best sample a smaller gap on all 15 instances at both sizes ($P=6.1\times10^{-5}$; all uncorrected, post hoc, and outside the test log). A circuit-sampled budget sweep at $n=12$ from 5\% to 45\% of the space (30 instances, eight ADMM iterations) shows the reversal at 10\% (read-out loss uncorrected, $P=0.033$), 18.75\%, and 45\%, with inconclusive nulls for the read-out at 5\% ($P=0.40$) and the best sample at 30\% ($P=0.21$); without the top-up, early termination of COBYLA left up to 29\% of the budget unspent.

These results concern the training budget we gave to the optimized QAOA: 23 optimizer evaluations per subproblem in the main comparison and 5-56 in the sweep. A longer optimization could concentrate the final state enough to pass \eqref{eq:readout}; we do not claim that read-out scoring always reverses a comparison. Per-instance optimization is in any case the weak link: every protocol that optimizes the angles from random or warm-started initial values is significantly worse than SA ($r\ge+0.918$, $P<10^{-10}$), and with independent draws optimized QAOA ($p=2$) remains far behind SA in mean gap ($10^3\gu$ of 36.8 against 1.96; Table~\ref{tab:baselines}).

\subsection{Transferred Angles Solve the ADMM Subproblems}
\label{subsec:transfer}

Transferred-angle QAOA ($p=2$ and $p=3$ alike) found the exact optimum of every subproblem on all 80 instances, 3.4 percentage points more than SA (95\% CI 1.8 to 5.3; $P=1.5\times10^{-5}$, 17 non-tied pairs), and is equivalent to the SDP within the lenient margin. Because it solves every subproblem exactly, its ADMM trajectory coincides with that of the exact solver. Training need not use the target size: angles trained at $n=6$ and $n=8$ solve 99.8\% and 99.9\% of the $n=12$ subproblems exactly, and the three angle sets differ by at most 0.07~rad in any component, consistent with parameter concentration. Angles trained at $n=12$ gave a mean $\gu$ of at most 0.0015 at every size from 6 to 16 (15 instances per size, eight ADMM iterations, budget 18.75\% of the search space, 25\% at $n=8$ and the whole space at $n=6$) and solved 99.5\% to 100\% of the subproblems at every penalty from 0.5 to 4. They also solved every recorded subproblem of the general family exactly.

\subsection{Stronger Classical Baselines and Cost}
\label{subsec:baselines}

\begin{figure*}[t!]
\centering
\includegraphics[width=\textwidth]{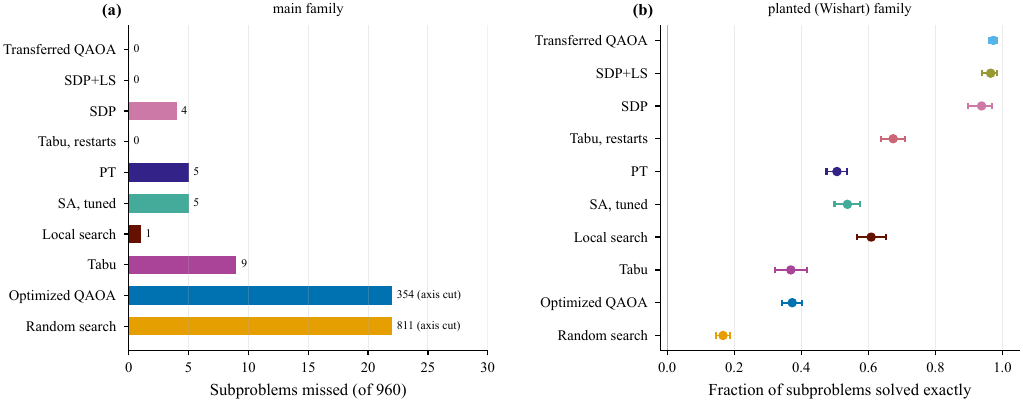}
\caption{Subproblem accuracy on the identical subproblems of the exact-solver ADMM trajectory ($n=12$, 80 instances, 960 subproblems per solver, $\B=768$). (a) Main family: number of the 960 subproblems that each solver did not solve exactly (the axis is cut for optimized QAOA, 354, and random search, 811). (b) Planted (Wishart) family: fraction of subproblems solved exactly, instance-level mean, and 95\% bootstrap interval. Optimized QAOA is drawn by the exact-distribution sampler in both panels.}
\label{fig:solvers}
\end{figure*}

Table~\ref{tab:baselines} adds classical competitors stronger than default SA, including the three stronger heuristics. On the 960 subproblems of the exact-solver trajectory, transferred QAOA is equivalent at the instance level within 2 percentage points (outside the log) to each of seven classical solvers (tabu search with and without restarts, SDP+LS, the SDP, local search, PT, and SA tuned on the extended grid; Fig.~\ref{fig:solvers}(a)), and no difference in fraction solved exactly survives Holm correction.

\begin{table}[t]
\centering
\begin{threeparttable}
\caption{Stronger baselines ($n=12$, ADMM-embedded, 80 instances, $\B=768$, $\rho=2$). $10^3\gu$: mean normalized gap $\times10^3$; $P$: exact signed-rank test of transferred QAOA against the row method on $\gu$ (non-tied pairs); time: median wall-clock seconds per subproblem in the Kaggle CPU environment (serial Python implementations).}
\label{tab:baselines}
\setlength{\tabcolsep}{3pt}
\footnotesize
\begin{tabular}{@{}lcccc@{}}
\toprule
Method & $10^3\gu$ & $\Pr[\text{opt}]$ & $P$ vs transferred & Time (s) \\
\midrule
Transferred $p=2$ & 0.00 & 1.000 & n/a & 0.0025\tnote{a} \\
Tabu, restarts & 0.00 & 1.000 & 1.0 (0) & 0.0018 \\
SDP+LS & 0.00 & 1.000 & 1.0 (0) & 0.016 \\
Local search & 0.03 & 0.999 & 1.0 (1) & 0.0012 \\
SDP & 0.08 & 0.994 & 0.13 (4)\tnote{b} & 0.015 \\
PT & 0.25 & 0.993 & 0.13 (4) & 0.0070 \\
SA, tuned (small grid) & 0.44 & 0.993 & 0.031 (6) & 0.011 \\
Tabu & 0.60 & 0.990 & $7.8\times10^{-3}$ (8) & 0.0017 \\
SA, restarts & 0.61 & 0.983 & $2.0\times10^{-3}$ (10) & 0.011 \\
SA & 1.96 & 0.966 & $1.5\times10^{-5}$ (17) & 0.011 \\
QAOA $p=2$, optimized\tnote{c} & 50.1 & 0.552 & n/a & 0.071\tnote{a} \\
Random & 81.2 & 0.177 & $<10^{-10}$ (80) & 0.0002 \\
\bottomrule
\end{tabular}
\begin{tablenotes}[para]\footnotesize
\item[a] Classical simulation of the circuit; a device needs 0.25 to 0.27~s for 768 shots. \item[b] Descriptive, not part of the test log. \item[c] Circuit sampler; with independent draws 36.8 and 0.641 (Section~\ref{subsec:sampling}).
\end{tablenotes}
\end{threeparttable}
\end{table}

Evaluations are not equally expensive: the classical solvers other than random search took 1.16 to 16.4~ms per subproblem (including the relaxation for the SDP-based ones), whereas 768 shots took 0.253 to 0.274~s at the device's $3.30\times10^{-4}$ to $3.56\times10^{-4}$~s per shot (Section~\ref{sec:setup}), so a 768-shot subproblem takes 15 to 237 times longer (Table~\ref{tab:baselines}, note a), before compilation and job latency.

\subsection{Scaling at a Fixed Budget}
\label{subsec:fixedbudget}

\begin{figure*}[t!]
\centering
\includegraphics[width=\textwidth]{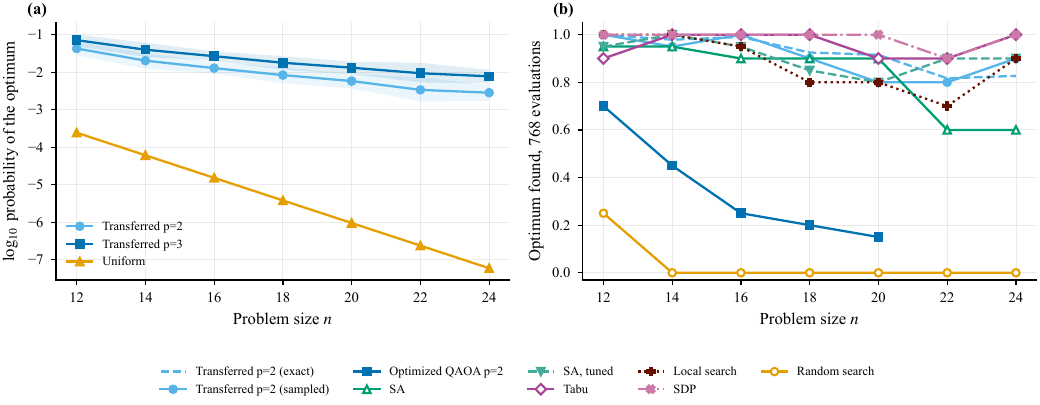}
\caption{Fixed budget of 768 evaluations, $n=12$ to $24$. (a) Probability of the optimum under the transferred $p=2$ and $p=3$ states and under uniform sampling, mean of $\log_{10}q$ over instances with 95\% interval. (b) Fraction of instances on which the optimum was found: exact value for the transferred $p=2$ state (dashed, no markers) and one 768-evaluation run per solver (markers).}
\label{fig:fixed}
\end{figure*}

With the budget held at 768 evaluations as $n$ grows from 12 to 24, where it covers $4.6\times10^{-5}$ of the space (Fig.~\ref{fig:fixed}; second ADMM subproblem; 20 instances per size, 10 at $n=22$ and 24), the probability $q$ of the optimum under the transferred state falls as $2^{-0.33n}$ for $p=2$ and $2^{-0.27n}$ for $p=3$ (least-squares fits over instances), while $2^{-n}$ falls by a factor of 4,096; the exact success probability at 768 samples stays, in the mean over instances, at or above 0.816 ($p=2$) and 0.966 ($p=3$). Growing amplification over uniform sampling \cite{golden2023numerical} is nevertheless a low bar on this family: tabu search and the SDP found the optimum on 90\% to 100\% of instances at every size, against 70\% falling to 15\% at $n=20$ for per-instance optimized QAOA (exact-distribution sampler), and no comparison of transferred QAOA with tuned SA or tabu search was significant ($P\ge0.31$, at most five non-tied pairs). Section~\ref{subsec:tts} therefore measures cost at budgets small enough not to saturate.

\subsection{A Planted Family With a Rugged Landscape}
\label{subsec:hard}

Landscape diagnostics of the second ADMM subproblem and of $F$ (medians over 80 instances at $n=12$ and 20 at $n=16$) show that the main family is easy: its subproblems have 4 single-flip local minima at $n=12$ and 7.5 at $n=16$, steepest descent from a random start reaches the optimum from 67\% and 53\% of all starting points, and at $n=12$ the Euclidean norm of the linear term $h$ is 0.79 times the Frobenius norm of the off-diagonal couplings. The planted subproblems have 45 and 204.5 local minima, basins of 6.0\% and 2.1\%, and a norm ratio of 0.021 at $n=12$, so the couplings set the landscape.

Table~\ref{tab:hard} and Fig.~\ref{fig:solvers} repeat the comparisons on 80 planted instances. On the identical subproblems of the exact-solver trajectory (960 per solver), transferred QAOA, with angles trained on held-out planted instances, found the exact optimum on 97.3\% of them; SDP+LS (96.5\%) and the SDP (93.8\%) did not differ significantly from it (cluster-adjusted McNemar $P=0.66$ and 0.071), whereas every local heuristic was far behind (the four strongest: tabu search with restarts 67.4\%, local search 60.8\%, tuned SA 53.8\%, PT 50.6\%; single-trajectory tabu search 36.9\%; all $P\le5\times10^{-6}$). Embedded in ADMM at $\rho=2$, the ordering is the same (97.4\%, 88.8\%, 79.8\%, and 48.8\% to 67.8\% for the four strongest local heuristics), and transferred QAOA had a smaller normalized gap than every local heuristic ($P<10^{-10}$) and the SDP ($P=2.9\times10^{-3}$); against SDP+LS it was smaller at $\rho=1$ ($P=4.3\times10^{-3}$) but not at $\rho=2$ ($P=0.40$). The planted-family angles were trained at the test size; angles retrained at $n=16$ fit about as well but differ by at least 0.52~rad in one component, even after accounting for the symmetries of the QAOA output distribution, so neither transfer across sizes nor parameter concentration is established on this family.

\emph{Near-degenerate competitors.} The Ising part of a planted subproblem is invariant under complementing all bits, and the symmetry-breaking field is weak, so the exact-optimum criterion also tests whether a solver resolves a few nearly degenerate low-energy states; for QAOA, whose mixer and initial state share the symmetry, only the cost layer can resolve them. We therefore re-scored the shared-trajectory runs from their logged costs by exhaustive enumeration of each subproblem, without rerunning any solver. The complement of the optimum, among the four best bitstrings on 724 of the 960 subproblems (median $\gu$ over all 960: 0.063), accounted for 22\% to 33\% of the strongest local heuristics' misses. These heuristics succeeded on 64.2\% to 74.6\% of subproblems when it was credited, against 98.2\% for transferred QAOA and 96.8\% for SDP+LS; on 90.4\% to 94.4\% when any of the four best bitstrings was credited, against 100\% and 98.8\%; and on 95.7\% to 98.2\% when every bitstring with $\gu\le0.1$ was credited, against 100\%. Transferred QAOA stayed ahead of each under every criterion (instance-level signed-rank $P\le4.9\times10^{-4}$, post hoc and outside the test log) and in mean gap at $\rho=2$ (Table~\ref{tab:hard}), so its advantage is not an artifact of the criterion. That advantage, however, lies in resolving the best of a handful of nearly degenerate states: under the loosest criterion, the local heuristics missed only 1.8\% to 4.3\% of the subproblems.

\begin{table}[t]
\centering
\begin{threeparttable}
\caption{Planted family ($n=12$, 80 instances, $\B=768$). $\Pr[\text{opt}]$: fraction of subproblems solved exactly, ADMM-embedded (12 iterations) and on the identical subproblems of the exact-solver trajectory (960 per solver); $10^3\gu$: mean normalized gap $\times10^3$; problem optimum: fraction of instances on which the best feasible iterate is the global optimum, embedded in ADMM and from a direct solution of $F$ with $K\B=9{,}216$ evaluations.}
\label{tab:hard}
\footnotesize\setlength{\tabcolsep}{1.5pt}
\begin{tabular}{@{}lccccccc@{}}
\toprule
 & \multicolumn{3}{c}{Solved exactly} & $10^3\gu$ & \multicolumn{3}{c}{Problem optimum} \\
\cmidrule(lr){2-4}\cmidrule(l){6-8}
Method & $\rho=2$ & $\rho=1$ & Shared & $\rho=2$ & $\rho=1$ & $\rho=2$ & Direct \\
\midrule
Transf.\ $p=2$ & 0.974 & 0.973 & 0.973 & 1.2 & 1.000 & 0.938 & 1.000 \\
Transf.\ $p=3$\tnote{a} & 0.976 & 0.973 & n/a & 1.1 & 1.000 & 0.938 & 1.000 \\
Transf.\ $p=2$, main\tnote{b} & 0.875 & 0.859 & n/a & 5.5 & 1.000 & 0.975 & n/a \\
SDP+LS & 0.888 & 0.806 & 0.965 & 3.2 & 0.962 & 0.875 & 0.912 \\
SDP & 0.798 & 0.747 & 0.938 & 7.4 & 0.888 & 0.750 & 0.788 \\
Tabu, restarts & 0.678 & 0.649 & 0.674 & 17.4 & 1.000 & 0.975 & 1.000 \\
Local search & 0.623 & 0.582 & 0.608 & 18.7 & 1.000 & 0.950 & 1.000 \\
SA, tuned (ext.) & 0.541 & 0.505 & 0.538 & 23.0 & 1.000 & 1.000 & 0.975 \\
PT & 0.488 & 0.468 & 0.506 & 26.2 & 1.000 & 0.988 & 0.988 \\
QAOA $p=2$, optim. & 0.407 & 0.371 & 0.373 & 42.2 & 0.975 & 0.962 & n/a \\
SA & 0.362 & 0.329 & n/a & 38.1 & 0.975 & 1.000 & n/a \\
Tabu & 0.353 & 0.290 & 0.369 & 40.9 & 0.962 & 0.938 & 0.262 \\
Random & 0.155 & 0.165 & 0.167 & 60.0 & 0.888 & 0.875 & 0.862 \\
Exact & 1.000 & 1.000 & 1.000 & 0.0 & 1.000 & 0.912 & n/a \\
\bottomrule
\end{tabular}
\begin{tablenotes}[para]\footnotesize\item[a] Training at $p=3$ converged to $\beta_1\approx0$, which merges the first two cost layers and reproduces the $p=2$ circuit (training $\gbar=0.377$ at both depths); this row carries no information about depth. \item[b] Transferred QAOA ($p=2$) with the main-family angles applied unchanged. Transf.: transferred; optim.: optimized.\end{tablenotes}
\end{threeparttable}
\end{table}

The subproblem advantage did not carry over to the ADMM level. At $\rho=1$, transferred QAOA, the exact solver, and the four strongest local heuristics found the problem optimum on all 80 instances. At $\rho=2$, where the exact solver itself missed it on 8.8\% of instances, transferred QAOA found it on 93.8\% and the strongest local heuristics on 95.0\% to 100\%; no logged ADMM-level comparison of transferred QAOA with tabu search with restarts, PT, tuned SA, or SDP+LS was significant ($P\ge0.0625$). By Theorem~\ref{thm:admm}(a), every iteration after the first at $\rho=1$ poses the same QUBO $F$, so an ADMM run gives a solver $K-1$ attempts at $F$.

\subsection{Time to Solution}
\label{subsec:tts}

\begin{figure*}[t!]
\centering
\includegraphics[width=\textwidth]{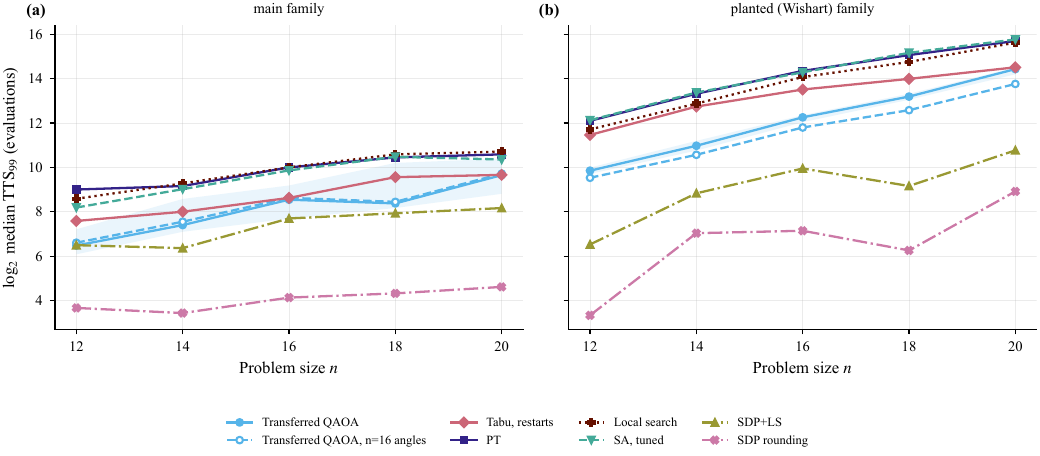}
\caption{Time to solution on the second ADMM subproblem, $n=12$ to 20, 20 test instances per size, for (a) the main and (b) the planted family: $\log_2$ of the median $\mathrm{TTS}_{99}$ in evaluations; the band shows the 95\% bootstrap interval for transferred QAOA. Run budgets of the classical solvers were chosen on held-out instances; SDP-based solvers exclude the relaxation; random search ($\mathrm{TTS}_{99}\propto2^{n}$) is omitted.}
\label{fig:tts}
\end{figure*}

Because a single budget of 768 evaluations saturates the classical solvers on the main family, Table~\ref{tab:tts} and Fig.~\ref{fig:tts} report $\mathrm{TTS}_{99}$ on the second ADMM subproblem from $n=12$ to 20 (20 test instances per size and family). Transferred QAOA is scored by its exact probability of the optimum in noiseless simulation, with angles trained at $n=12$ and, separately, with angles retrained at $n=16$. No local-heuristic or QAOA estimate was censored, and on the planted family, no local heuristic succeeded in every run on more than 10\% of instances. Tuning the classical solvers' other parameters at $n=12$ and keeping them at larger $n$ favors QAOA in the comparison of exponents.

\begin{table*}[t]
\centering
\begin{threeparttable}
\caption{Time to solution $\mathrm{TTS}_{99}$ (noiseless simulation for QAOA, $p=2$) on the second ADMM subproblem (median over 20 test instances), the exponent $\kappa$ in $\mathrm{TTS}_{99}\propto2^{\kappa n}$ fitted to the medians over $n=12$ to 20 with 95\% bootstrap interval (an interval reaching $\infty$ contains resamples with a censored median), and the planted-family $\mathrm{TTS}_{99}$ in seconds.}
\label{tab:tts}
\setlength{\tabcolsep}{4.2pt}
\footnotesize
\begin{tabular}{@{}lcccccccc@{}}
\toprule
 & \multicolumn{3}{c}{Main family, evaluations} & \multicolumn{3}{c}{Planted family, evaluations} & \multicolumn{2}{c}{Planted, seconds\tnote{a}} \\
\cmidrule(lr){2-4}\cmidrule(lr){5-7}\cmidrule(l){8-9}
Method & $n=12$ & $n=20$ & $\kappa$ & $n=12$ & $n=20$ & $\kappa$ & $n=12$ & $n=20$ \\
\midrule
Transferred $p=2$, angles from $n=12$ & 89 & 803 & 0.37 [0.25, 0.48] & 923 & 22,001 & 0.57 [0.54, 0.59] & 0.33 & 7.8 \\
Transferred $p=2$, angles from $n=16$ & 97 & 841 & 0.36 [0.25, 0.46] & 739 & 13,845 & 0.52 [0.49, 0.56] & 0.26 & 4.9 \\
Tabu, restarts\tnote{b} & 192 & 809 & 0.29 [0.21, 0.35] & 2,817 & 23,264 & 0.37 [0.31, 0.43] & 0.0061 & 0.050 \\
SA, tuned\tnote{c} & 292 & 1,308 & 0.29 [0.19, 0.40] & 4,466 & 55,588 & 0.45 [0.39, 0.50] & 0.059 & 1.1 \\
PT\tnote{b} & 512 & 1,536 & 0.22 [0.16, 0.32] & 4,378 & 52,950 & 0.45 [0.38, 0.49] & 0.022 & 0.41 \\
Local search\tnote{b} & 384 & 1,673 & 0.28 [0.23, 0.36] & 3,346 & 50,289 & 0.48 [0.44, 0.53] & 0.0049 & 0.068 \\
SDP+LS\tnote{d} & 90 & 287 & 0.25 [0.16, 0.32] & 92 & 1,741 & 0.44 [0.30, 0.75] & 0.031 & 0.11 \\
SDP rounding\tnote{d} & 13 & 24 & 0.14 [0.06, 0.24] & 10 & 483 & 0.52 [0.05, $\infty$] & 0.031 & 0.11 \\
Random & 18,860 & $4.8\times10^6$ & 1.00 & 18,860 & $4.8\times10^6$ & 1.00 & n/a & n/a \\
\bottomrule
\end{tabular}
\begin{tablenotes}[para]\footnotesize
\item[a] QAOA: device time per shot measured on \texttt{ibm\_marrakesh} at $n=12$ ($3.56\times10^{-4}$~s), applied to the noiseless $\mathrm{TTS}_{99}$; classical solvers: measured serial time per evaluation of our Python implementations in the Kaggle CPU environment. \item[b] Succeeded in every run on 10\% to 60\% of main-family instances at the selected budget, so the main-family values are upper bounds. \item[c] One anneal per run, temperatures of the tuned configuration. \item[d] Relaxation excluded from the evaluation count, included in seconds; censored on up to 25\% (SDP+LS) and 35\% (SDP rounding) of planted instances at some sizes; SDP+LS also succeeded in every run on up to 40\% of planted instances, so its planted values are upper bounds as well.
\end{tablenotes}
\end{threeparttable}
\end{table*}

On the planted family, transferred QAOA needed 923 evaluations at $n=12$, three to five times fewer than the local heuristics ($P=1.9\times10^{-6}$ against each, the smallest exact $P$ with 20 pairs), but its cost grew faster, as $2^{0.57n}$ against $2^{0.37n}$ to $2^{0.48n}$: the paired bootstrap places the excess of its exponent above zero against every local heuristic (by 0.08 to 0.20, $P\le0.002$), and at $n=20$ tabu search with restarts needed about as many evaluations ($P=0.28$). Angles retrained at $n=16$ lowered the exponent to 0.52 and the cost at $n=20$ but did not remove the excess against tabu search with restarts, PT, and tuned SA. These values count only the exact optimum; the near-degenerate competitors of Section~\ref{subsec:hard} were not recorded here. The prefactor also depends on the evaluation convention in Section~\ref{sec:design}: tabu search and local search incur $n$ evaluations per scan, and counting a scan as one evaluation would reduce their counts by up to a factor of $n$ and could reverse the prefactor in their favor. SDP+LS was the fastest classical solver with a run length on both families; on the planted family, it needed 10 and 13 times fewer evaluations than transferred QAOA at $n=12$ and 20, but its median was not monotone in $n$ and was censored on up to 25\% of instances, so its exponent could not be distinguished from that of QAOA. On the main family, the medians of three local heuristics are upper bounds (Table~\ref{tab:tts}, note b), so neither the prefactor nor the exponents are resolved there, and we draw no conclusion from them.

The evaluation count also hides the cost of an evaluation: a device shot took $3.56\times10^{-4}$~s, a classical evaluation $6.34\times10^{-7}$ to $2.05\times10^{-5}$~s. In seconds and before any hardware error, transferring QAOA on the planted family took 54 and 157 times longer than tabu search with restarts at $n=12$ and 20, and 11 and 72 times longer than SDP+LS, including its relaxation. In noiseless simulation at $p=2$, transferred QAOA thus has, on the planted family and under our evaluation convention, a prefactor advantage over local heuristics in evaluations, a larger exponent, and no advantage over SDP-based solvers or in time. This is a statement about $p=2$: the decay of the probability of the optimum was slower at $p=3$ on the main family and slows with depth in other fixed-angle studies \cite{boulebnane2024solving,montanezbarrera2025toward}.

\subsection{ADMM-Level Behavior and the Penalty}
\label{subsec:admm}
\label{subsec:penalty}

At the benchmark's penalty $\rho=2$, subproblem accuracy and the ADMM outcome diverge. With exact, SDP, or transferred-angle subproblem solutions, the mean primal residual falls to $3\times10^{-5}$ after 12 iterations, yet the best feasible iterate is the global optimum of \eqref{eq:mbo} on only half of the instances (SA: 55\%); optimized QAOA and random search, whose noisy trajectories keep changing $z$, find it on 76\% and 75\% (optimized QAOA with independent draws: 78.8\%). In the exact-solver runs, the residual of 79 of the 80 instances is divided by exactly $1+\rho=3$ at each of the last six iterations, the signature of Theorem~\ref{thm:admm}(b) for a binary iterate that no longer changes; by part (c), that iterate is a stalling point, which at $\rho=2$ can be suboptimal. Optimized QAOA's advantage over the exact solver is therefore not a quantum effect; random search, also noisy, matches it.

\begin{figure*}[t!]
\centering
\includegraphics[width=\textwidth]{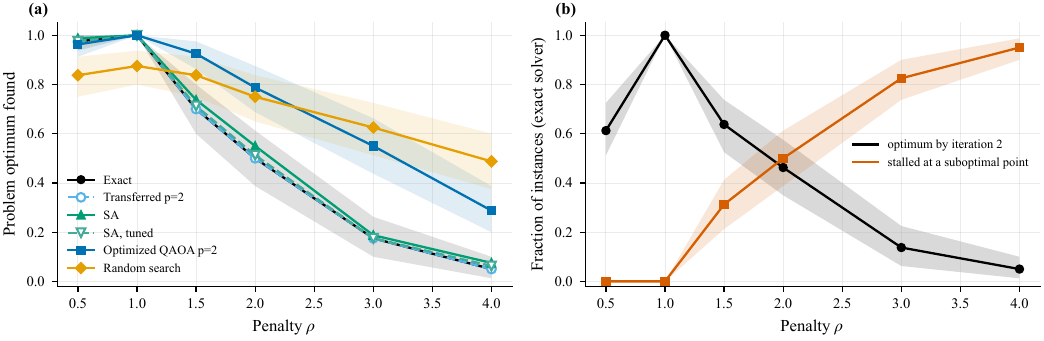}
\caption{Penalty sweep ($n=12$, 80 instances, 12 iterations; QAOA from the exact-distribution sampler). (a) Fraction of instances on which ADMM's best feasible iterate is the global optimum of \eqref{eq:mbo}; the exact and transferred-angle curves coincide, and tuned SA and SA lie within one and four instances of them, respectively, at every penalty. (b) Exact solver: fraction of instances that reached the global optimum by the second iteration and that stalled at a suboptimal bitstring (held over the last three iterations). Means with 95\% bootstrap intervals.}
\label{fig:penalty}
\end{figure*}

Theorem~\ref{thm:admm} predicts that this advantage comes from the penalty. We repeated the embedded comparison on the same 80 instances at $\rho\in\{0.5,1,1.5,2,3,4\}$ (Fig.~\ref{fig:penalty}). At $\rho=1$, exact subproblem solutions reached the global optimum by the second iteration on all 80 instances, as part (a) states; SA, tuned SA, transferred QAOA, and optimized QAOA also found it on every instance, random search on 87.5\%, and every solver did better than at $\rho=2$ ($P\le2.0\times10^{-3}$ for each). For $\rho>1$, the exact solver stalled at a suboptimal bitstring increasingly often (31.2\%, 50\%, 82.5\%, and 95\% of instances at $\rho=1.5$, 2, 3, and 4), whereas noisy solvers degraded more slowly. As parts (b) and (c) state, every such stall satisfied \eqref{eq:fp}, and the residual ratio equaled $1+\rho$ to within $10^{-7}$ whenever the bitstring was held. At $\rho=0.5$, no instance stalled at a suboptimal bitstring. For problem \eqref{eq:mbo}, the ADMM-level advantage of noisy solvers at $\rho=2$ is therefore an artifact of the penalty.

\emph{Solving the reduced QUBO directly.} The natural reference is one solution of $F$ with the budget of a whole ADMM run, $K\B=9{,}216$ evaluations. On the main family, the direct solution differed from ADMM at $\rho=1$ on at most one instance for every solver except random search, and with every accurate solver it beat ADMM at $\rho=2$ by 38 to 40 instances; on this family, ADMM at $\rho=1$ thus amounts to a direct solution of $F$, and ADMM at $\rho=2$ did worse. On the planted family, the direct solution found the optimum on every instance with transferred QAOA, local search, and tabu search with restarts, and the decomposition acted mainly as a restart mechanism for single-trajectory tabu search (26.2\% directly, 96.2\% inside ADMM at $\rho=1$) and the SDP.

\subsection{Mixed-Binary Families That Do Not Reduce to a QUBO}
\label{subsec:general}

To test what survives without the structure of Theorem~\ref{thm:admm}, we generated 80 instances of
\begin{equation}
\min_{x,z}\ \tfrac12x^{\top}Hx-c^{\top}x+z^{\top}Qz\ \ \text{s.t.}\ Ax+Bz=b,\ |x_i|\le1,
\label{eq:general}
\end{equation}
with $z\in\{0,1\}^{12}$, $x\in\mathbb{R}^{16}$, eight coupling constraints, a random Gaussian $A$, and $H=LL^{\top}+0.5I$ with Gaussian $L$. ADMM's $x$-update becomes a box-constrained convex quadratic program, the $z$-update is still a QUBO, and neither Theorem~\ref{thm:admm} nor its extension to a general quadratic applies. We obtain the problem optimum from a box-constrained quadratic program for each feasible bitstring. The instances use seeds 10000 to 10079; of the first 40, 36 admit a feasible $x$ for every $z$ and all 40 admit one for at least 80\% of the bitstrings, every ADMM iterate was feasible, and the box constraint is active at the continuous optimum for a median of 45\% of bitstrings.

\begin{figure*}[t!]
\centering
\includegraphics[width=\textwidth]{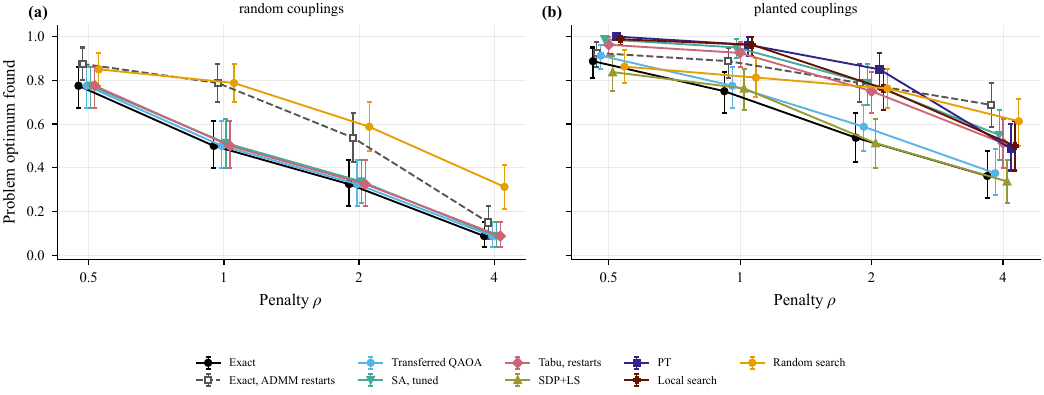}
\caption{General mixed-binary family \eqref{eq:general} (12 ADMM iterations, 80 instances per panel) with (a) random couplings drawn as in the main family and (b) Wishart planted couplings. Fraction of instances on which the best feasible iterate is the problem optimum, against $\rho$, for each subproblem solver and for ADMM with four restarts and exact subproblem solutions (dashed); mean and 95\% bootstrap interval. In (a), the exact, transferred-angle, tabu-with-restarts, and tuned-SA curves nearly coincide.}
\label{fig:general}
\end{figure*}

Transferred angles reproduced the exact solver's best feasible solution on every instance at every penalty and solved exactly every subproblem whose accuracy was recorded (the first 40 instances, $\rho=1$ and 2). On those 40 instances, optimized QAOA (exact-distribution sampler) found the problem optimum on 85\% and 57.5\% at $\rho=1$ and 2, as often as random search, against 45\% and 32.5\% for exact solutions. At the ADMM level (Fig.~\ref{fig:general}(a)), however, $\rho=1$ is no longer special, and accurate solvers do worse as the penalty grows. On all 80 instances, exact subproblem solutions found the problem optimum on 77.5\%, 50\%, 32.5\%, and 8.8\% at $\rho=0.5$, 1, 2, and 4, and tabu search with restarts and tuned SA matched them to within one instance; random search found it on 85\%, 78.8\%, 58.8\%, and 31.2\% and beat transferred QAOA at $\rho=1$, 2, and 4 ($P\le6.8\times10^{-4}$) but not at $\rho=0.5$ ($P=0.59$). Restarts, the classical remedy for nonconvex ADMM \cite{takapoui2020simple}, recover much of this advantage without a randomized subproblem solver: four exact-solver runs of three iterations from random continuous starting points found it on 87.5\%, 78.8\%, 53.8\%, and 15.0\%, and transferred QAOA was worse than this restarted exact ADMM at $\rho=1$, 2, and 4 ($P\le5.0\times10^{-5}$).

With Wishart planted couplings (80 instances, seed offset 40000, $B$ from $[-0.1,0.1]$, planted-family angles; Fig.~\ref{fig:general}(b)), the subproblem ranking carried over (at $\rho=1$, transferred QAOA solved 97.8\% of subproblems exactly and SDP+LS 96.1\%, against 64.9\% to 75.3\% for the four strongest local heuristics), but the ADMM ranking reversed: at $\rho=1$, the problem optimum was found on 77.5\% of instances with transferred QAOA, 75.0\% with exact solutions, and 76.2\% with SDP+LS, against 92.5\% to 96.2\% with the local heuristics and 88.8\% with restarted exact ADMM ($P\le1.6\times10^{-3}$ at $\rho=1$ for transferred QAOA against each; at $\rho=2$ and 4, $r$ from $+0.70$ to $+0.94$; at $\rho=0.5$ no difference survived Holm correction). Accurate subproblem solutions stall, whereas errors from less-accurate heuristics move $z$.

\subsection{Verification and Statistical Accounting}
\label{subsec:verification}

The correctness gate passed at every level. We disclose two defects that benchmark results alone did not reveal: a Hamiltonian construction that halved the linear terms contributed by the couplings, which the gate now excludes, and the shared random stream of Section~\ref{subsec:sampling}. Of the 392 logged tests (157 confirmatory, 235 exploratory), 159 survive the global Bonferroni threshold ($P<1.28\times10^{-4}$) and 233 remain significant after Holm correction within their families; the survivors include every best-sample comparison of a QAOA protocol with random search in the main setting (the budget control in the opposite direction) and every subproblem-level comparison of transferred QAOA with a local heuristic on the planted family at 768 evaluations and in time to solution up to $n=18$ (and $n=20$ except against tabu search with restarts). Of the 392 tests, 96 (50 confirmatory, 23 survivors) come from the mechanism, optimizer-ablation, Hamiltonian-defect, CVaR and initialization, size, noise, budget, coupling-density, and MaxCut studies, which this paper reports only where stated. No paired slope difference survives the global threshold, which its 4,000 bootstrap resamples cannot reach ($P\ge1/4001$); all are exploratory, and several are not significant even nominally.
\section{Discussion}
\label{sec:discussion}

\subsection{Scoring and Budget Are Design Decisions}
Best-sample scoring asks what a decomposition method asks of its subproblem solver and is how every classical baseline is implicitly scored \cite{lykov2023sampling}; read-out scoring raises the bar for QAOA by a factor of up to $\B/R$ (Proposition~\ref{prop:scoring}). QAOA benchmarks should log every sampled bitstring, state the scoring rule and $\B/R$, report the evaluations actually used, draw the shots of every evaluation independently and test that they are, and report cost in seconds as well as in evaluations, because a matched number of evaluations hides a cost ratio of 17 to 560 between a device shot ($3.56\times10^{-4}$~s) and a classical evaluation in our implementations (Section~\ref{subsec:tts}).

\subsection{What Transferred Angles Achieve}
That transferred angles work on moving ADMM subproblems is consistent with parameter concentration \cite{brandao2018forfixed,akshay2021concentration,wurtz2021fixed}, but on the benign main family, seven classical solvers also solve essentially every subproblem. On the planted family, the advantage over the local heuristics is real but bounded: it lies in resolving a handful of nearly degenerate low-energy states, is a prefactor with a larger exponent at $p=2$, and does not extend to SDP-based solvers or wall-clock time. The main open question, as in decay-rate comparisons of fixed-angle QAOA on other problems \cite{dupont2022calibrating,boulebnane2024solving,montanezbarrera2025toward}, is whether greater depth, angles trained for the probability of the optimum, or larger sizes change this comparison. The known limits of bounded-depth QAOA against local algorithms \cite{farhi2020needs,hastings2019classical,barak2022classical} concern approximation ratios, not time to solution.

\subsection{The Penalty Decides What the Subproblem Is}
For problem \eqref{eq:mbo}, the binary step is a proximal step on the reduced QUBO; at $\rho=1$ it is the reduced QUBO itself, and for $\rho>1$ exact solutions stall at suboptimal points increasingly often. That a noisy subproblem solver yields better ADMM-level solutions than an exact one at $\rho=2$ is therefore a statement about the penalty, not about the value of noise; the exploration that helps is also provided by random search or by restarts \cite{takapoui2020simple}. In the general families, where no penalty removes the proximal term, the same holds for $\rho\ge1$. Studies of QAOA inside ADMM should report $\rho$, the ADMM-level outcome alongside subproblem quality, random-search and restart controls, and, when the problem reduces to a QUBO, a direct solution of it. The three-block heuristic of \cite{gambella2020multiblock}, which Qiskit Optimization implements with a penalty that starts at $\rho=10^4$ and grows during the run \cite{qiskitoptimization}, is outside Theorem~\ref{thm:admm}; whether its schedule causes stalling of the same kind is open, and these controls would answer it.

These results do not demonstrate a quantum advantage, but they indicate where one would have to come from: a family on which QAOA's exponent, not only its prefactor, beats the strongest classical solvers, SDP-based ones included, in time rather than evaluations, on hardware that preserves the probability of good solutions, and inside a decomposition that does not simply repeat the reduced problem.

\section{Limitations}
\label{sec:limitations}

\emph{Scope.} We studied the two-block ADMM scheme with a fixed penalty on a main problem class that reduces to a QUBO; only the two synthetic general families need a decomposition, and we did not run the three-block, adaptive-penalty heuristic of \cite{gambella2020multiblock}. All instances are synthetic; the planted families come from a single ensemble parameter ($\alpha=0.5$), and their near-symmetry under complementation makes the exact-optimum rate partly a test of resolving a weak field. Most experiments use budgets that are a sizable fraction of the search space; the fixed budget extends to $n=24$, and the time to solution extends to $n=20$ on the second ADMM subproblem only.

\emph{Sampling.} The circuit sampler reused one random stream per subproblem (Section~\ref{subsec:sampling}). Only the main optimized-QAOA arm at $\rho=2$ has an independent-draw replicate, so the values of the other optimized variants in Fig.~\ref{fig:main}, of the 40-instance shared-trajectory comparison, and of the budget sweep carry this caveat. We did not repeat these runs with the exact-distribution sampler. The scoring comparisons that support our conclusions, at $n=12$, 16, and 20 and at every penalty, used independent draws.

\emph{Depth, training, and cost.} Transferred QAOA was scored in noiseless simulation at $p=2$ (also $p=3$ in several experiments, including the fixed budget) with angles trained on the mean gap, so the exponents are statements about $p=2$ over five sizes in a pre-asymptotic range. In a depolarizing-noise check (12 instances, five ADMM iterations, probability $p_1$ on single-qubit and $2p_1$ on two-qubit gates), transferred angles kept a mean gap below 0.001 up to $p_1=1\%$ and reached 0.036 at 5\% (random search: 0.078); hardware noise is outside the scope of this paper. Per-instance optimization used at most 56 optimizer evaluations per subproblem. Classical parameters other than the run budget were tuned at $n=12$; run budgets were selected on 20 held-out instances per size; and the conversion to seconds used serial Python implementations and the device time per shot at $n=12$, choices that favor QAOA.

\emph{Statistics.} The planted-family and stronger-heuristic experiments were designed after the main-family results.

\section{Conclusion}
\label{sec:conclusion}

In every family we tested, QAOA, as the binary-subproblem solver of a two-block ADMM, did not give a better ADMM outcome than the best classical choice of solver and penalty. It was competitive at the subproblem level only when run with transferred angles and credited with its best sample, as classical solvers are; with 23 optimizer evaluations, read-out scoring reversed the comparison. At depth 2, it then matched seven classical solvers on the main family and, in noiseless simulation, beat the local heuristics on a rugged planted family by resolving nearly degenerate states and by a prefactor in evaluations, but with a larger exponent; it stayed behind SDP rounding with local search, and far behind in seconds. For problem \eqref{eq:mbo}, ADMM re-solves the reduced problem at $\rho=1$, and exact solutions increasingly stall for $\rho>1$; without this structure, less accurate solvers or restarts improved the ADMM outcome for $\rho\ge1$. The benchmark protocol lets the same controls be applied to deeper circuits, application-derived instances, and decomposition schemes used in practice.

\section*{Data and Code Availability}
The code, notebooks, per-instance results, test log, and analysis scripts are available at \url{https://github.com/sanim-yousuf/qaoa-admm-benchmark}. 

\section*{Appendix\\Proofs of the Analytical Results}
\label{sup:proofs}
Throughout, $\|\cdot\|$ is the Euclidean norm, $w=(\rho-1)/2$ is the weight of the proximal term in Theorem~\ref{thm:admm}, and line numbers in the proof of Theorem~\ref{thm:admm} refer to Algorithm~\ref{alg:admm}.

\begin{proof}[Proof of Proposition~\ref{prop:scoring}]
\emph{Idea:} compare the probabilities that the two methods miss the good set $\Omega$ altogether.

\emph{(a)} The read-out shots are among the logged shots (Algorithm~\ref{alg:qaoa}), so the best logged bitstring is at least as good as the best read-out bitstring.

\emph{(b)} The $R$ read-out shots are independent draws from the final state, and random search makes $\B$ independent uniform draws. Hence
\begin{align*}
\Pr[\text{read-out misses } \Omega]&=(1-\pi)^R,\\
\Pr[\text{random search misses } \Omega]&=(1-u)^{\B}.
\end{align*}
The read-out does at least as well if and only if $(1-\pi)^R\le(1-u)^{\B}$; taking the $R$-th root gives $\pi\ge1-(1-u)^{\B/R}$, which is \eqref{eq:readout}. To size this threshold, let $t=(\B/R)\,u\le1$, and note that $\B/R\ge1$ because the read-out is part of the budget. Two elementary inequalities give
\begin{equation*}
1-e^{-t}\ \le\ 1-(1-u)^{\B/R}\ \le\ t.
\end{equation*}
The right inequality is Bernoulli's inequality, $(1-u)^s\ge1-su$ for $s\ge1$; the left one uses $1-u\le e^{-u}$. Finally, $1-e^{-t}\ge(1-e^{-1})\,t\approx0.63\,t$ on $[0,1]$, because the concave function $1-e^{-t}$ lies above the chord joining its values at $t=0$ and $t=1$.

\emph{(c)} If every shot is scored, the read-out is the whole budget; set $R=\B$ in (b).
\end{proof}

\begin{proof}[Proof of Proposition~\ref{prop:fixed}]
\emph{Idea:} the shots are independent, so the optimum is missed only if every shot misses it.

Each shot misses the optimal set with probability $1-q$, so all $\B$ shots miss it with probability $(1-q)^{\B}$, and $1-q\le e^{-q}$ gives \eqref{eq:miss}. The best sample has gap 0 when the optimum is found, and at most $G$ otherwise, so its expected gap is at most $G\,(1-q)^{\B}$. Finally, $e^{-q\B}\le\epsilon$ if and only if $q\ge\ln(1/\epsilon)/\B$.
\end{proof}

\begin{proof}[Proof of Theorem~\ref{thm:admm}]
\emph{Idea:} after the first iteration, the dual variable is fixed by the continuous iterate. Substituting this one identity into the QUBO gives the proximal form, and parts (b) and (c) then follow from how the anchor $a^k$ moves.

\emph{Step 1 (dual identity).} Line~5 minimizes the augmented Lagrangian over $x$, so, as in the standard analysis of nonconvex ADMM \cite{li2015global},
\begin{equation*}
x^{k+1}-c+\lambda^k+\rho\,(x^{k+1}+Bz^{k+1}-b)=0.
\end{equation*}
Line~6 sets $\lambda^{k+1}=\lambda^k+\rho\,(x^{k+1}+Bz^{k+1}-b)$, which by this equation is
\begin{equation}
\lambda^{k+1}=c-x^{k+1}\qquad\text{for every }k\ge0.
\label{eq:dualid}
\end{equation}

\emph{Step 2 (proximal form).} Applying \eqref{eq:dualid} with $k-1$ in place of $k$ gives $\lambda^k=c-x^k$ for every $k\ge1$, which turns the linear term \eqref{eq:zqubo} into $h^k=B^{\top}c+(\rho-1)B^{\top}x^k-\rho B^{\top}b$. Expanding \eqref{eq:reduced} gives, up to a constant, $F(z)=z^{\top}Qz+\tfrac12\|Bz\|^2+(c-b)^{\top}Bz$. Subtracting $F(z)$ from the QUBO objective leaves, up to a constant,
\begin{equation*}
w\|Bz\|^2-2w\,(a^k)^{\top}Bz=w\|Bz-a^k\|^2-w\|a^k\|^2,
\end{equation*}
with $a^k=b-x^k$. The last term does not depend on $z$, which proves \eqref{eq:prox}. At $\rho=1$, the weight $w$ is zero, so line~4 minimizes $F$ from the second iteration on. Because $x=b-Bz$ eliminates the constraint of \eqref{eq:mbo}, the minimizers of $F$ are exactly the $z$-parts of the global optima of \eqref{eq:mbo}; with $x=b-Bz$, the bitstring returned at the second iteration therefore gives a global optimum, which Algorithm~\ref{alg:admm} returns as its best feasible iterate. This proves (a).

\emph{Step 3 (the anchor moves toward $B\bar z$).} Suppose line~4 returns $\bar z$ at an iteration $k\ge1$, so $z^{k+1}=\bar z$, and let $\bar x=b-B\bar z$. Inserting $\lambda^k=c-x^k$ (valid because $k\ge1$) into line~5 gives $(1+\rho)(x^{k+1}-\bar x)=x^k-\bar x$, which in terms of the anchor reads
\begin{equation}
a^{k+1}-B\bar z=\frac{a^k-B\bar z}{1+\rho}.
\label{eq:anchor}
\end{equation}
Each iteration that returns $\bar z$ moves the anchor by the fraction $\rho/(1+\rho)$ of the remaining distance toward $B\bar z$.

\emph{Step 4 (part (b)).} When $z^{k+1}=\bar z$, the residual of line~7 is $\delta^{k+1}=\|x^{k+1}-\bar x\|=\|a^{k+1}-B\bar z\|$. If also $z^k=\bar z$, that is, $\bar z$ was returned at iteration $k-1$, then $\delta^k=\|x^k-\bar x\|=\|a^k-B\bar z\|$, and \eqref{eq:anchor} gives $\delta^{k+1}=\delta^k/(1+\rho)$, whether or not $\bar z$ is optimal.

\emph{Step 5 (part (c), only if).} Write
\begin{equation*}
\Phi_j(z)=F(z)+w\|Bz-a^j\|^2
\end{equation*}
for the objective at iteration $j$. If $\bar z$ is returned at every iteration from $k$ on, then, because line~4 is solved exactly, $\Phi_j(z)\ge\Phi_j(\bar z)$ for all $z$ and all $j\ge k$; and \eqref{eq:anchor}, which applies at every $j\ge k$, gives $a^j\to B\bar z$. Letting $j\to\infty$ yields
\begin{equation*}
F(z)+w\|B(z-\bar z)\|^2\ \ge\ F(\bar z)\quad\text{for all } z,
\end{equation*}
which is \eqref{eq:fp}.

\emph{Step 6 (part (c), if).} Fix $z$ and let $D(a)$ be the advantage of $\bar z$ over $z$ when the anchor is $a$, that is, the objective at $z$ minus the objective at $\bar z$. The term $w\|a\|^2$ cancels, so $D$ is an affine function of $a$. It is nonnegative at $a=a^k$, because the exact solver returned $\bar z$ at iteration $k$, and at $a=B\bar z$, by \eqref{eq:fp}. By \eqref{eq:anchor}, $a^{k+1}$ lies on the segment between these two points, so $D(a^{k+1})\ge0$. Hence $\bar z$ is again a minimizer and, with ties broken in its favor, is returned at iteration $k+1$. Repeating the argument at iterations $k+1,k+2,\dots$ shows that $\bar z$ is returned at every later iteration.

\emph{Step 7 (remaining statements).} At $\rho=1$, \eqref{eq:fp} reads $F(\bar z)\le F(z)$ for all $z$. For $\rho>1$ the term $w\|B(z-\bar z)\|^2$ is nonnegative, so every global minimizer $z^\star$ satisfies \eqref{eq:fp}, and setting $z=z^\star$ in \eqref{eq:fp} gives $F(\bar z)-F(z^\star)\le w\|B(z^\star-\bar z)\|^2$. The right side of \eqref{eq:fp} does not decrease as $\rho$ grows, so a stalling point for one penalty remains one for every larger penalty. Steps 5 and 6 do not use the sign of $w$, so the characterization holds for every $\rho>0$. For $\rho<1$, $w<0$, and setting $z=z^\star$ in \eqref{eq:fp} gives $F(\bar z)\le F(z^\star)-|w|\,\|B(z^\star-\bar z)\|^2\le F(z^\star)$, so $\bar z$ is a global minimizer, and \eqref{eq:fp} is then the sharpness condition stated in the theorem.
\end{proof}

\begin{proof}[Proof of Remark~\ref{rem:general}]
\emph{Idea:} repeat Steps 1, 2, 3, and 5 of the proof of Theorem~\ref{thm:admm} with the weight $\rho I-H$ in place of $(\rho-1)I$. With continuous part $\tfrac12x^{\top}Hx-c^{\top}x$, the optimality condition of the $x$-update (line~5 with $H$ in place of $I$) reads $Hx^{k+1}-c+\lambda^k+\rho\,(x^{k+1}+Bz^{k+1}-b)=0$, so line~6 gives $\lambda^{k+1}=c-Hx^{k+1}$, and hence $\lambda^k=c-Hx^k$ for $k\ge1$. With $a^k=b-x^k$, the objective of line~4 is, up to a constant, $z^{\top}Qz+(\lambda^k)^{\top}Bz+\tfrac{\rho}{2}\|Bz-a^k\|^2$, and the reduced objective is $F(z)=\tfrac12(b-Bz)^{\top}H(b-Bz)-c^{\top}(b-Bz)+z^{\top}Qz$. Substituting $\lambda^k=c-Hb+Ha^k$ and subtracting $F(z)$ leaves, up to a constant,
\begin{align*}
&(a^k)^{\top}HBz-\tfrac12\|Bz\|_H^2+\tfrac{\rho}{2}\|Bz-a^k\|^2\\
&\qquad=\tfrac12\|Bz-a^k\|^2_{\rho I-H}+\tfrac12\|a^k\|_H^2,
\end{align*}
and the last term does not depend on $z$. For $H=I$, this is the proximal term of \eqref{eq:prox}. If line~4 returns $\bar z$ at an iteration $k\ge1$ and $\bar x=b-B\bar z$, line~5 now gives $(H+\rho I)(x^{k+1}-\bar x)=H(x^k-\bar x)$, that is, $a^{k+1}-B\bar z=(H+\rho I)^{-1}H\,(a^k-B\bar z)$. This matrix contraction reduces to \eqref{eq:anchor} for $H=I$. Its eigenvalues $\eta_i/(\eta_i+\rho)$, with $\eta_i>0$ the eigenvalues of $H$, lie in $(0,1)$, so the anchor still converges to $B\bar z$ and Step 5 applies with $\Phi_j(z)=F(z)+\tfrac12\|Bz-a^j\|^2_{\rho I-H}$, giving $F(z)+\tfrac12\|B(z-\bar z)\|^2_{\rho I-H}\ge F(\bar z)$ for all $z$. In general, however, the contraction does not keep $a^{k+1}$ on the segment between $a^k$ and $B\bar z$, and the residual is not divided by a single scalar, so Steps 4 and 6 do not apply. For $H=\eta I$, the weight $\rho I-H$ vanishes at $\rho=\eta$, where line~4 minimizes $F$ from the second iteration on, as in Step 2.
\end{proof}

\section*{Acknowledgment}
We acknowledge the use of IBM Quantum services for this work. The views expressed are those of the authors and do not reflect the official policy or position of IBM or the IBM Quantum team. The authors used Claude (Anthropic), an AI assistant, to help draft and edit the text of all sections, including the appendix, to draw the schematic diagram of Fig.~\ref{fig:method}, and to help write part of the analysis code. The authors designed the study, ran all experiments and all code, checked every reported number against the computed outputs, and take full responsibility for the content.
\bibliographystyle{IEEEtran}
\bibliography{references}

\end{document}